\documentclass[12pt]{article}
\usepackage{hyperref}
\usepackage{amsmath}
\usepackage{epsfig}
\usepackage{amssymb}
\usepackage[numbers,sort&compress]{natbib}

\def \sect #1 {\setcounter{equation} 0\section{#1}}

\def \nn {\nonumber}
\def \be  {\begin{equation}}
\def \ee  {\end{equation}}
\def \bea  {\begin{eqnarray}}
\def \eea  {\end{eqnarray}}
\def \ba {\begin{eqnarray*}}
\def \ea {\end{eqnarray*}}
\def \bb  {\begin{thebibliography}}
\def \eb  {\end{thebibliography}}

\def \lab #1 {\label{#1}}

\def \e {\mbox{e}}

\def \fracs #1#2 {\mbox{\small $\frac{#1}{#2}$}}

\def \bin #1#2 {{\left({#1}\atop{#2}\right)}}
\def \as {\relax\ifmmode\alpha_s\else{$\alpha_s${ }}\fi}

\def \al #1 {\frac {\as({#1})}{\pi} }
\def \ds #1 {\ooalign{$\hfil/\hfil$\crcr$#1$}}

\def \d {{\rm d}}

\def\hepph  #1 {{\tt hep-ph/#1}}

\begin{document}

{\Large
\begin{center}
{\bf Power Corrections to Electroweak Boson Production

\medskip

from Threshold Resummation}

\vspace{5mm}

\end{center}
}

\par \vspace{2mm}
\begin{center}
{\bf George Sterman${}^{\,a}$,}
\hskip .2cm
{\bf Werner Vogelsang${}^{\,b}$  }\\[5mm]
\vspace{5mm}
${}^{a}\,$ C.N.\ Yang Institute for Theoretical Physics and Department of Physics and Astronomy\\
Stony Brook University, Stony Brook, 
New York 11794 -- 3840, U.S.A.\\
${}^{b}\,$ Institute for Theoretical Physics, T\"ubingen University, 
Auf der Morgenstelle 14, \\ 72076 T\"ubingen, Germany\\[2mm]
\end{center}

\vspace{5mm}

\begin{abstract}
We study the power corrections for electroweak boson  production that are implied by threshold resummation, which
we have extended to massive particles produced at measured transverse momentum, $p_T$, and
rapidity.  Power corrections in the resulting expressions arise
from ambiguities in the low-scale behavior of the
perturbative running coupling.   
Arguing for the relevance of the eikonal approximation, we show that such power corrections begin
at order $1/p_T^2$ in full QCD, consistent with fixed-order,  massive-gluon analysis.  For large-$N$ Mellin moments,
the leading behavior is $N^2/p_T^2$, which exponentiates
along with the logarithms of threshold resummation.

\end{abstract}

\newpage

\tableofcontents

\newpage
\section{Introduction}

Leading-power, factorized cross sections are central to the experimental program at high-energy accelerators, but the need for precision 
control over Standard Model calculations motivates the study of power-suppressed corrections in these cross sections.   
In this paper, we study the production of electroweak bosons, $\beta=\gamma,\gamma^*,W^\pm,Z$, with large transverse momenta, $p_T$,
in hadronic collisions.   For an observed color-singlet particle of momentum $p$ 
produced in collisions between hadrons $A$ and $B$, the generic, leading-power, factorized form is
\bea
p^0\, \frac{ d\sigma_{AB\to \beta+X} }{d^3p} \ &=&\
\sum_{ab} \int dx_a dx_b\ \phi_{a/A}(x_a,\mu^2)\, \phi_{b/B}(x_b,\mu^2)\ \omega_{ab\to \beta+X} 
\left(x_ap_A,x_bp_B,p,\mu,\alpha_s(\mu^2)\right)
\nn\\[2mm]
&\ & \hspace{5mm} +\ {\rm power\ corrections}\, ,
\label{eq:fact-leading-power}
\eea
where the sum is over parton flavors with distributions $\phi_{a/A}$ and $\phi_{b/B}$, and where 
$\mu$ denotes a factorization/renormalization scale.  
For real photons, we neglect fragmentation contributions, and for heavy bosons, we take $p_T\sim m$, with $m$ the mass of the produced particle, to avoid the presence of large ratios.  Corrections to Eq.\ (\ref{eq:fact-leading-power}) are suppressed 
relative to the leading term 
by powers in $p_T$, or, generically, a hard scale $Q$, and it is these corrections that we study below.  

The nature of power corrections to factorized cross sections of the form of Eq.\ (\ref{eq:fact-leading-power}) was discussed  in Refs.\ \cite{Qiu:1990xy,Qiu:1991wg}, where arguments were presented that at fixed orders in perturbation theory, power corrections up to $1/Q^2$ in 
the hard scale $Q$ take the form of factorized  twist-four matrix elements.  This leaves, however, the possibility of power corrections associated with resummations of higher-order leading-power corrections, especially near ``partonic threshold", defined as the integration region where
\bea
\hat s \ = \ x_ax_bS\ \rightarrow \ Q^2
\label{eq:partonic-threshold}
\eea
 in Eq.\ (\ref{eq:fact-leading-power}).  
 (We note that for single-particle cross sections, the hard scale $Q$ is in general a function of the four-vectors $x_aP_A$, $x_bP_B$ and $p$.)
 Near partonic threshold, real radiation is suppressed, and although the hard scattering function 
 $\omega_{ab\to\beta+X}$ remains infrared finite order-by-order in perturbation theory, the cancellation holds only up to scales that  vanish when  
Eq.\ (\ref{eq:partonic-threshold}) becomes an equality.   The threshold-resummed cross section generates a divergent series in this limit, which should be reinterpreted
as implying the presence of power corrections to the factorized cross section 
\cite{Korchemsky:1994is,Beneke:1995pq,Beneke:1996xy,Catani:1996yz,Forte:2006mi}.    
As was shown in Refs.~\cite{Laenen:2000hs,Laenen:2000ij}, this approach also predicts that power corrections to the inclusive
Drell-Yan cross section start at $1/Q^2$ in full QCD.   Extensions of these ideas to the case of high-$p_T$ direct-photon production
for integrated rapidity were 
carried out in Ref.~\cite{Sterman:2004yk}, where leading power corrections of order $1/p_T^2$  to the cross section were anticipated,
relying on the formalism of ``joint resummation" \cite{Kulesza:2002rh}, which organizes threshold and transverse-momentum resummations.
The analysis we give below generalizes and extends the treatment of inclusive Drell-Yan in \cite{Laenen:2000hs,Laenen:2000ij} and that
of prompt photons in~\cite{Sterman:2004yk}, including 
systematic treatments 
of final-state soft radiation at measured rapidity for the observed particle,
and of the role of the recoiling colored parton in $AB\to \beta X$. 

Our discussion is also inspired by recent arguments for the absence of nonperturbative $1/Q$ corrections in the hard scale for a phenomenologically-relevant class of event shapes \cite{Luisoni:2020efy} in lepton pair annihilation and for the hadronic production of color neutral bosons at large transverse momentum \cite{FerrarioRavasio:2020guj,Caola:2021kzt}.   The extensive arguments given to support this conclusion 
rely in part on a much older treatment of Drell-Yan cross sections in hadronic collisions, 
based on renormalon analysis in 
abelian theories with large numbers of flavor in Refs.\ \cite{Beneke:1995pq} and \cite{Beneke:1996xy}.  Related analyses rely on identifying sensitivity to a finite gluon mass with a nonperturbative representation of the running coupling 
\cite{Dasgupta:1996hh,Gardi:1999dq,Dasgupta:1999zm},
in the ``dispersive" representation of the QCD running coupling at low scales \cite{Dokshitzer:1995qm}.  

Again, hard-scattering functions  $\omega_{ab\to \beta+X}$ in Drell-Yan are singular at ``partonic threshold", where the incoming partons have just enough energy to produce the observed final state.   At each order in 
the strong coupling $\alpha_s$, the functions $\omega_{ab\to \beta+X}$ are characterized by singular plus distributions whose arguments vanish at partonic threshold.   
For example, in the inclusive
Drell-Yan cross section the relevant argument is $1-z$, with $z=Q^2/(x_ax_bS)$  and the 
leading distributions are of the form
\bea
\alpha_s^n\, \left [ \frac{ \ln^{2n-1}(1-z)}{1-z} \right ]_+\, .
\eea
These singular distributions, in turn, can be controlled by threshold resummation, which organizes all such logarithms, including those associated with the running of the coupling. The argument of the running coupling vanishes as $z$ approaches 1.
We will encounter similar behavior below, extending 
the analysis for photons in Ref.\ \cite{Sterman:2000pt} to massive electroweak bosons produced at high $p_T$.

The relevant transform for the inclusive production of color-neutral particles involves, 
as we review below, the Mellin or Laplace factor with moment variable, $N$, 
\bea
(1-\beta_r\cdot k/Q)^N\ \sim \ \exp [ -N\beta_r\cdot k /Q ]\ \left( 1 +\ {\cal O}(1/N)\right )\, ,
\eea
where $k^\mu$ is the sum of all radiation from a set of two incoming and one outgoing Wilson line, the latter characterized by velocity $\beta_r$, which is in the direction of recoil 
to the observed photon 
or color neutral boson at partonic threshold. The denominator that sets the scale of the exponent, $Q$, is the hard scale of the scattering, of order $p_T$.   As $N$ increases, real radiation in the moments is restricted to energies of order $Q/N$, which can be of the order of $\Lambda_{\rm QCD}$.  

We will carry out our analysis using the eikonal approximation, which we will argue is appropriate for both real and virtual radiation at these low scales (see, for example, Refs.\ \cite{Kidonakis:1998bk} and \cite{Becher:2007ty}).   
In the discussion that follows, we show that in full QCD, it is possible to infer the presence of specific power corrections to color-neutral boson production at large 
$p_T\sim Q$.   These corrections appear through (exponentiating) terms of the form
\bea
 \left (\frac{N}{Q} \right)^n
\int_0^\kappa {d\mu}\,\mu^{n-1}\; A_i \left( \alpha_s(\mu^2) \right)\, ,
\label{eq:F-gamma}
\eea
for an integer $n$, 
with $A_i(\alpha_s)$ a flavor-dependent function of the strong coupling, and $\kappa$ an arbitrary cutoff.
Such integrals are ambiguous in perturbation theory because of the behavior of the perturbative running coupling.   If re-expressed as an inverse Borel transform, the singularity of the running coupling is mapped onto a singularity on the positive real axis in the Borel plane \cite{Korchemsky:1994is}.
In particular, we show that  threshold resummation implies the presence of such terms beginning
at $n=2$ for these cross sections, consistent with the results of Ref.\ \cite{FerrarioRavasio:2020guj}.   

In the following section, we review the double moment method of \cite{Sterman:2000pt}
 for particle production at high $p_T$ and measured rapidity,
\footnote{The use of double moments in threshold resummation was introduced in \cite{Catani:1989ne}.
For a related treatment of resummation at measured rapidity, see Ref.\ \cite{Westmark:2017uig,Banerjee:2017cfc}.}
extending the formalism to the massive case.   Of particular interest is the expression for the cross
section as an inverse Mellin (or Laplace) transform, from which we infer the exponential suppression of nonperturbative
behavior in the large-moment region.   Section 3 reviews the construction of the cross section for
color-singlet production in the eikonal 
approximation, found from matrix elements of Wilson lines, which will generate 
all corrections that behave as powers of $N/p_T$, with $N$ the Mellin moment.  We go on to show that final-state interactions 
cancel in these cross sections.   Section 4 reviews salient features of an explicit all-orders resummation 
for these processes, in terms of webs  \cite{Sterman:1981jc,Gatheral:1983cz,Gardi:2010rn,Laenen:1998qw,Berger:2003zh}.
We present results generated by the lowest-order web, verifying the absence of linear power corrections at this level.   
We show, in Sec.\ 5, how general features of exponentiated webs
derived in Sec.\ 4 lead to an expression
for the resummed exponent, and we verify the absence of linear powers of $N/p_T$   to all orders in the web expansion.
We briefly discuss further applications and and conclude.

\section{Double-moment cross sections}

To set the stage for our analysis of logarithms associated with partonic threshold, we discuss the factorized
cross section for the production of an electroweak boson as a function of transverse momentum and rapidity. 
In particular, we will present the moment-space 
formulation of the cross section, which is an essential ingredient of our method for organizing the threshold 
logarithms and will subsequently allow us to use the eikonal approximation to identify power corrections associated with renormalons.
The following derivations follow those in our analysis of prompt photon production~\cite{Sterman:2000pt}; however, we extend them here
to the case of a massive boson. Although we have mostly in mind the Drell-Yan process and the $W$ and $Z$ electroweak bosons, 
the application to Higgs production is immediate.
 
At measured transverse momentum $p_T$ and rapidity $\eta$, the factorized form for the cross section
for $AB\to \beta X$, as in Eq.\ (\ref{eq:fact-leading-power}), is given by
\bea
\label{eq:1}
\frac{ p_T^3  d\sigma_{AB\to \beta X}}{dp_Td\eta} &=& \sum_{a,b}\, 
\int_{-\frac{U}{S+T-m^2}}^1 dx_a \, \phi_{a/A}\left(x_a,\mu^2\right) \,
\int_{\frac{-x_a (T-m^2)-m^2}{x_aS+U-m^2}}^1 dx_b \, \phi_{b/B}\left(x_b,\mu^2\right) \, 
\frac{p_T^3 d\hat{\sigma}_{ab}}{{dp_Td\eta}}\nn\\[2mm]
&\equiv&\sum_{a,b}
\int_{-\frac{U}{S+T-m^2}}^1 dx_a \, \phi_{a/A}\left(x_a,\mu^2\right) \,
\int_{\frac{-x_a (T-m^2)-m^2}{x_aS+U-m^2}}^1 dx_b \, \phi_{b/B}\left(x_b,\mu^2\right)
\omega_{ab}\left(\hat{x}_T,\hat\eta,r,\frac{\mu^2}{\hat{s}}\right) \, , \nn\\
\eea
where the $\omega_{ab}$ are dimensionless hard-scattering functions. 
We have introduced a set of kinematic variables: The mass of the produced boson 
$\beta$ is denoted by $m$.
The hadronic Mandelstam variables are $S=(P_A+P_B)^2$, $T=(P_A-p)^2$, $U=(P_B-p)^2$,
with $p$ the boson four-momentum. In terms of the boson's transverse momentum, mass and 
rapidity we have
\bea
T&=&m^2-\sqrt{S}\sqrt{m^2+p_T^2}\,{\mathrm{e}}^{-\eta}\,,\nn\\[2mm]
U&=&m^2-\sqrt{S}\sqrt{m^2+p_T^2}\,{\mathrm{e}}^{\eta}\,.
\eea
For future reference we also define
\be
x_T\,\equiv\,\frac{p_T+\sqrt{m^2+p_T^2}}{\sqrt{S}}\,\equiv\,\frac{p_T+m_T}{\sqrt{S}}\,.
\ee
As indicated, the partonic hard-scattering functions may be written as functions of 
\bea
\hat{x}_T&\equiv& \frac{x_T}{\sqrt{x_ax_b}} \,,\nn\\[2mm]
\hat\eta&\equiv&\eta-\frac{1}{2}\ln \frac{x_a}{x_b}\,,\nn\\[2mm]
r &\equiv&\frac{p_T}{m_T}\,,
\label{eq:variables}
\eea
and $\mu^2/\hat{s}$, with $\hat{s}=x_ax_bS$.

We now take combined Mellin/Fourier moments of the hadronic cross section at fixed $p_T$,
\be
\Sigma_{AB\to \beta X}(N,M,p_T)\,\equiv\,\int_{-\infty}^\infty d\eta \,{\mathrm{e}}^{i M \eta}\int_0^{x_{T,\mathrm{max}}^2}
dx_T^2 \,\big( x_T^2\big)^{N-1}\,\frac{p_T^3 d\sigma_{AB\to \beta X}}{dp_Td\eta}\,.
\ee
Here $x_{T,\mathrm{max}}^2$ is the kinematic upper limit on $x_T^2$, given at fixed rapidity by
\be
x_{T,\mathrm{max}}^2\,=\,\frac{\cosh^2 \eta}{(1-r)^2}\left(1-\sqrt{1-\frac{1-r^2}{\cosh^2 \eta}}\right)^2\,.
\label{eq:xt-max}
\ee
Applying the moment integrals to the expression in the second line of Eq.~(\ref{eq:1}) we
find, after some straightforward algebra:
\bea\label{momdef}
\Sigma_{AB\to \beta X}(N,M,p_T)&=&\sum_{a,b}\int_0^1dx_a\,x_a^{N+i M/2} \phi_{a/A}\left(x_a,\mu^2\right)   \int_0^1dx_b\,x_b^{N-i M/2}
\phi_{b/B}\left(x_b,\mu^2\right)\nn\\[2mm]
&\times& \int_{-\infty}^\infty d\hat{\eta}\,{\mathrm{e}}^{i M \hat{\eta}}\int_0^{\hat{x}_{T,\mathrm{max}}^2}
d\hat{x}_T^2\,\left( \hat{x}_T^2\right)^{N-1}\,\omega_{ab}\left(\hat{x}_T,\hat\eta,r,\frac{\mu^2}{\hat{s}} \right)\nn\\[2mm]
&\equiv&\sum_{a,b} \tilde{\phi}_{a/A}^{N+1+\frac{i M}{2}} (\mu^2)  \, \tilde{\phi}_{b/B}^{N+1-\frac{i M}{2}} (\mu^2)   \,
\tilde{\omega}_{ab}\left(N,M,r,\frac{\mu^2}{p_T^2}\right)\,.
\eea
Here the Mellin moments of the parton distribution functions are defined as usual by
\be
\tilde{\phi}_{i/H}^n\left(\mu^2\right)\,\equiv\,\int_0^1 dx \, x^{n-1}\,\phi_{i/H}(x,\mu^2)\,,
\ee
and we have introduced
\be\label{dobmom}
\tilde{\omega}_{ab}\left(N,M,r,\frac{\mu^2}{p_T^2}\right)\,\equiv\,
\int_{-\infty}^\infty d\hat{\eta}\,{\mathrm{e}}^{i M \hat{\eta}}\int_0^{\hat{x}_{T,\mathrm{max}}^2}
d\hat{x}_T^2\,\left( \hat{x}_T^2\right)^{N-1}\,\omega_{ab}\left(\hat{x}_T,\hat\eta,r,\frac{\mu^2}{\hat{s}}\right)\,,
\ee
with 
\be
\hat{x}_{T,\mathrm{max}}^2\,=\,\frac{\cosh^2 \hat\eta}{(1-r)^2}\left(1-\sqrt{1-\frac{1-r^2}{\cosh^2 \hat\eta}}\right)^2\,.
\ee
As seen in Eq.~(\ref{eq:xt-max}), in moment space the contribution to the cross section for a given 
partonic channel factorizes into a product of the moments of the parton distributions and a double moment
of the partonic hard-scattering function. We note that in the limit $m \to 0$ (or $r\to 1$) Eq.~(\ref{momdef}) 
reverts to the corresponding expression for the case of prompt photons discussed in Ref.~\cite{Sterman:2000pt}. 

In threshold resummation, we organize terms that become singular at partonic threshold, when the active partons
$a$ and $b$ have just enough energy to produce a boson at fixed $\hat{x}_T$ and $\hat \eta$.
In order to investigate the moment-space expression near threshold, it is convenient to introduce the 
variable
\be
\zeta\,\equiv\,\frac{s_4}{\hat{s}}\,\equiv\,\frac{\hat{s}+\hat{t}+\hat{u}-m^2}{\hat{s}}\,,
\label{eq:zeta-def}
\ee
where 
\bea
\hat{t}&=&m^2+x_a (T-m^2)\,,\nn\\[2mm]
\hat{u}&=&m^2+x_b (U-m^2)\,.
\eea
The invariant $s_4$ provides a natural measure of the distance from threshold. 
In terms of $\hat{x}_T, \hat\eta$ and $r$ we have
\be\label{eq:zetadef}
\zeta\,=\,1+\hat{x}_T^2\,\frac{1-r}{1+r}-\frac{2\hat{x}_T}{1+r}\,\cosh\hat{\eta}\,,
\ee
which may be inverted to give
\be
\hat x_T(\zeta) \,=\,  \frac{ \cosh\hat{\eta}}{ 1-r } \left(1 \, -\, \sqrt{ 1 -\frac{(1-r^2)(1-\zeta)}{\cosh^2 \hat \eta}}  \,\right)\,.
\ee
Using this to replace the integration over $\hat{x}_T^2$ in~(\ref{dobmom}) we have for our double moments
\bea
\tilde{\omega}_{ab}\left(N,M,r,\frac{\mu^2}{p_T^2}\right)&=&\int_{-\infty}^\infty d\hat{\eta}\,{\mathrm{e}}^{i M \hat{\eta}}\,\frac{1+r}{(1-r)^{2N-1}}\,
\left(\cosh\hat{\eta}\right)^{2N-2}\int_0^1 d\zeta\,
\frac{\left(1-\sqrt{1-\frac{(1-r^2)(1-\zeta)}{\cosh^2\hat\eta}}\,\right)^{2N-1}}{\sqrt{1-\frac{(1-r^2)(1-\zeta)}{\cosh^2\hat\eta}}}\nn\\[2mm]
&\times&\omega_{ab}\left(\hat{x}_T,\hat\eta,r,\frac{\mu^2}{\hat{s}}\right)\nn\\[2mm]
&=& \int_{-\infty}^\infty d\hat{\eta}\,{\mathrm{e}}^{i M \hat{\eta}}\,\frac{1+r}{\cosh \hat \eta} \,
\left(  \frac{\cosh \hat{\eta}}{{(1-r)}}\left(1-\sqrt{1-\frac{1-r^2}{\cosh^2 \hat{\eta}}}\,\right) \right)^{2N-1}
\nonumber\\[2mm]
&\times& \int_0^1 \frac{d\zeta}{\sqrt{1-\frac{(1-r^2)(1-\zeta)}{\cosh^2\hat\eta}}}\,
\left( \frac{1-\sqrt{1-\frac{(1-r^2)(1-\zeta)}{\cosh^2 \hat \eta}} }{  1-\sqrt{1-\frac{1-r^2}{\cosh^2 \hat{\eta}}}  } \right)^{2N-1}
\omega_{ab}\left(\hat{x}_T,\hat\eta,r,\frac{\mu^2}{\hat{s}}\right)\, .\nn\\
\label{eq:massive-moments}
\eea
The form given by the second equality shows that the $\zeta$ integrand is exponentially suppressed away from $\zeta =0$
at large $N$. We can therefore expand the fraction raised to the $(2N-1)$st power in the $\zeta$ integral to find
\bea
\label{eq:zeta-expand}
\tilde{\omega}_{ab}\left(N,M,r,\frac{\mu^2}{p_T^2}\right) &=& \int_{-\infty}^\infty d\hat{\eta}\,{\mathrm{e}}^{i M \hat{\eta}}\,\frac{1+r}{\cosh \hat \eta} \,
\left(  \frac{\cosh \hat{\eta}}{{(1-r)}}\left(1-\sqrt{1-\frac{1-r^2}{\cosh^2 \hat{\eta}}}\,\right) \right)^{2N-1}
\nonumber\\[2mm]
&&\hspace*{-3cm}\times\; \int_0^1 \frac{d\zeta}{\sqrt{1-\frac{(1-r^2)(1-\zeta)}{\cosh^2\hat \eta}}}\,
\exp \left[  - (2N-1)  \frac{\zeta}{2} \left ( 1 + \frac{ \cosh \hat \eta}{\sqrt{r^2 + \sinh^2\hat \eta }}\right) \right ] \omega_{ab}\left(\hat{x}_T,\hat\eta,r,\frac{\mu^2}{\hat{s}}\right) + \dots\,,\nn\\
\eea
with corrections that are suppressed by powers of $1/N$. In this form, we can carry out a ``standard" resummation of logarithms of $N$ for 
$\tilde{\omega}_{ab}$, although for nonzero mass $m$ the effective value of $N$ is linked in a mild way to $\hat \eta$.   
The $\hat \eta$ integrand still decreases rapidly for large $\hat \eta$, in the same manner as seen in the massless case~\cite{Sterman:2000pt}.

Given any determination of the large-$N$ behavior of $\tilde{\omega}_{ab}$,
the behavior of the physical cross section is given in turn by the inverse of the double transform,
\begin{eqnarray} \label{inverse} 
\frac{p_T^3 \d\sigma_{AB\to \beta X}}{\d p_T \d\eta}
&=&\frac{1}{2 \pi} \int_{-\infty}^{\infty} dM \, \e^{-i M \eta} \, 
\frac{1}{2 \pi i} \int_{{\cal C}} dN \, (x_T^2)^{-N} \,
\Sigma_{AB\to \beta X}(N,M,p_T) \,,
\end{eqnarray}
where ${\cal C}$ denotes a suitable contour in the complex-Mellin plane~\cite{Sterman:2000pt}, extending into the left half-plane where the integrand is exponentially suppressed.
For large values of the moment variable, that is, when $N\ge Q/\Lambda_{\rm QCD}$, the integrand in this expression is suppressed at least as $\exp [ \ln x_T (Q/\Lambda_{\rm QCD}) ]$.   
In our arguments below, we will identify nonperturbative corrections proportional to $(N/Q)^2$ in an exponentiated, resummed hard-scattering function.   
Our arguments, however, apply only when $N\Lambda_{\rm QCD} /Q$ remains order unity or smaller, so that the large-$N$ behavior of the physical cross section
converges as long as $x_T<1-\Lambda_{\rm QCD}/Q$.   
This is the case, for example, in the minimal resummation prescription \cite{Catani:1996yz}.

The outcome of this analysis is that in the physical cross section the inverse transform in $\hat x_T$ will remain exponentially suppressed at large $|N|$ along our inverse contour, as $\exp [ - N\ln (1/x_T) ]$.   This enables us to argue that the very large $N$ behavior  of the resummed exponent does not influence the result.   
That is, we only need to follow $N$ to order $p_T/\Lambda$.

\section{Eikonal cross sections}
\label{sec:eikonal}

Near partonic threshold, the limit $\zeta\to 0$ in Eq.\ (\ref{eq:zeta-def}),  the underlying processes for electroweak boson production are quark-antiquark annihilation to 
gluon plus boson, and (anti)quark-gluon to (anti)quark plus boson, accompanied by soft gluons, both of the form,
\bea
a\ +\ b \rightarrow \ \beta \ +\ r\, ,
\label{eq:abc-gamma}
\eea
where again, $\beta$ represents the color-singlet particle, and where we refer to $r$ as the recoil parton, which is taken here to 
be a massless quark, antiquark or gluon.     

The eikonal cross section is appropriate for soft radiation, and is naturally relevant for partonic threshold.   As we recall below, it is generated from cross sections involving products of
Wilson lines, which act as sources for gluons.   
We begin by discussing the kinematics at partonic threshold.
This enables us to show that to leading power in $N$, the transform
of the hard-scattering function in Eq.\ (\ref{eq:zeta-expand}) 
(in the general massive case) is determined by an eikonal cross section with a Laplace transform that depends on the soft radiation.  Here, and in the remainder of the paper, for brevity of notation, we will replace the value $2N-1$, which appears in Eq.\ (\ref{eq:zeta-expand}), by simply $N$. 

For large enough $N$, the moment-space hard scattering functions, 
$\tilde \omega_{ab}(N,M)$, are well-approxi\-ma\-ted by their eikonal approximations, in which terms quadratic in soft radiation are neglected compared to linear powers that are contracted with hard vectors, $k^2 \ll 2p\cdot k$.  
The contributions of corrections to the eikonal approximation start with terms 
that include relative factors of the general form $\beta_k^\mu |k_0|/Q$, with $k_0$ the energy of the radiation
and $\beta_k$ the velocity vector for $k$.   
Contributions of such terms to the variables $s_4$ in Eq.\ (\ref{eq:zeta-def}) and $\zeta$ in Eq.\ (\ref{eq:zetadef}),
which measure the distance to partonic threshold, are suppressed 
by a factor of $\beta_k\cdot \beta_r/N$ relative to the leading eikonal behavior.   In the special case of radiation in the recoil direction ($\beta_k = \beta_r$), the radiation of momentum $k^\mu$ will factor into a partonic jet function.   
Such jets contribute to the Laplace transform in Eq.\ (\ref{eq:zeta-expand}) through a factor 
$\exp[ -N m_{\rm jet}^2/Q^2]$, with nominal contributions like $N/Q^2$, rather than $N^2/Q^2$.   Taking as an assumption that $N^2/Q^2$ contributions will be larger than those of order $N/Q^2$, the eikonal approximation will give a good first approximation to the pattern of $1/Q^2$ power corrections.   This is the assumption we shall entertain, as motivation to work in eikonal approximation.

\subsection{The large-$N$ transform and threshold kinematics}

At large values of $N$ in Eq.\ (\ref{eq:zeta-expand}), the integral will be restricted to a region where $\zeta = s_4/\hat s$ is of order  $1/N$.   We would like to relate the integral over $\zeta$ to an integral over the 
momenta of final states.   In general, this is a nonlinear relation, because both $s_4$ and $\hat s$ depend on these momenta.   We can, however, relate $\hat s$ and $s_4$, and we find that for soft radiation, 
\bea
\hat s\ =\ \hat s_{\rm min} \ +\ {\cal O}(s_4)
\label{eq:hat-s-min}
\eea
where $\hat s_{\rm min}$,  the invariant corresponding to a final state consisting only of the color-singlet boson and the recoil parton, is 
\bea
\hat s_{\rm min}\ = \ \left( p_T^2 + m^2 \right)\, \cosh^2\hat \eta\, \left [ 1 + \sqrt{ 1 - \frac{1-r^2}{\cosh^2\hat\eta} } \right ]^2\, .
\label{eq:Qm-def}
\eea
As defined in Eq.\ (\ref{eq:variables}), $r=p_T/m_T$.  For the region of interest, we can represent the approximations that connect the $\zeta$ integral to a Laplace transform in $s_4/\hat s_{\rm min}$ as,
\bea
\int_0 d\zeta\, \left( 1 - \zeta\right)^N\ \rightarrow \int_0 d\zeta\, \ \e^{-N\zeta}\ \rightarrow\  \int_0 \frac{ds_4}{\hat s_{\rm min}}\, \exp \left( -\, N\frac{s_4}{\hat s_{\rm min}}\right)\, ,
\eea
where corrections are suppressed by powers of $s_4/\hat s \sim 1/N$.   As we have observed above, this is the level of accuracy of the eikonal cross section.  

At partonic threshold, the final state consists of only the color-singlet boson and the light-like recoil parton, $r$.   In the 
partonic center of mass, the spatial momentum of $p_r$ is back-to-back with the color-singlet particle, and depends 
on the rapidity $\hat \eta$ of that particle through
\bea
p_r^0\ &=&\ \sqrt{ p_T^2 \cosh^2\hat \eta + m^2 \sinh^2 \hat \eta }\, ,
\nn\\[2mm]
p_r^3\ &=&\ -\, \sqrt{ p_T^2 + m^2 } \, \sinh\hat \eta\, ,
\nn\\[2mm]
{\bf p}_{r,T} \ &=&\ -\, {\bf p}_T\, .
\label{eq:pr}
\eea
The explicit expression of $s_4$ for a state consisting of a recoil parton of momentum $p_r$ and soft radiation $k$ is
\bea
s_4\ =\ (k+p_r)^2\ =\ (k+ p_r^0\beta_r)^2\ \sim \ 2p_r^0\, \beta_r\cdot k\,,
\label{eq:s_4-beta_c}
\eea
where $\beta_r$ is a light-like velocity, with $\beta_r^0=1$, and where in the final relation we invoke the eikonal approximation.   In these terms, we will denote the transform function as
\bea
\exp \left( -\, N\frac{s_4}{\hat s_{\rm min}}\right)\ =\ \exp \left( - N\,\frac{ \beta_r\cdot k}{Q} \right )
\label{eq:transform-notation}
\eea
where the hard scale $Q$ is defined by
\bea
Q\ \equiv\ \frac{\hat s_{\rm min}}{2p_r^0}\, ,
\label{eq:Q-def-eikonal}
\eea
with $\hat s_{\rm min}$ given by (\ref{eq:Qm-def}) and $p_r^0$ by (\ref{eq:pr}).
We readily check that in the massless limit, $Q \rightarrow 2p_T\cosh\hat\eta$.   

In summary, we have shown that for large moment variable $N$, the total momentum of radiation $k$ in the hard-scattering function is
limited to regions where $s_4\sim p_r\cdot k < 1/N$.   In this region, up to corrections suppressed by
powers of $1/N$, the eikonal approximation holds and the $N$-dependence in the transform is approximated
by Eq.\ (\ref{eq:transform-notation}).

\subsection{Eikonal cross sections and Wilson lines}

We now turn to the construction of the eikonal approximation of the hard-scattering function 
$\tilde \omega_{ab}$ in Eq.\ (\ref{eq:zeta-expand}). For a generic color-singlet boson 
production process $ab\to\beta r$ we have \cite{Laenen:1998qw}
\bea
\tilde{\omega}^{\rm (eik)}_{abr} (N,Q,\hat{\eta},\mu)\ =\   H_{abr}(p_T,\hat{\eta},\mu) \times\, 
  \frac{\tilde \sigma^{\rm (eik)}_{abr} (N/Q,\hat{\eta},\mu,\epsilon)}{ \tilde \phi_{a/a}^{\rm (eik)}\left( N_a,\mu,\epsilon \right) 
\tilde \phi_{b/b}^{\rm (eik)}\left( N_b,\mu,\epsilon \right )}\, ,
\label{eq:tilde-omega}
\eea
where, again, $N$ stands for $2N-1$ in (\ref{eq:zeta-expand}), and $Q$, which is of order $p_T$, is defined  in Eq.\ (\ref{eq:Q-def-eikonal}).   
It will be convenient here and in the following to exhibit the label of the recoil parton, $r$, on the eikonal cross section and hard-scattering function.
The purely virtual short-distance function $H_{abr}$ is independent of $N$ and begins 
with the partonic Born cross section, which for convenience we normalize to the zeroth-order eikonal cross section (see below), 
\bea
H_{abr}(p_T,\hat{\eta},\mu)\ =\
\frac{1}{\sigma_{abr}^{{\rm (eik,0)}} }\frac{p_T^3 \d\hat{\sigma}^{\rm (Born)}_{ab\to  \beta r}}{\d p_T \d\hat{\eta}}\, \Big(1 + {\cal O}\left(\as(\mu^2)\right) \Big)\, .
\label{eq:Habr-def}
\eea
Dependence on $N$ is all in the eikonal cross section, $\tilde \sigma^{\rm (eik)}_{abr}$ and eikonal parton distributions $\tilde \phi^{\rm (eik)}_{i/i}$.   
As is characteristic of direct photon and similar single-particle cross sections, the moment values of the distributions are scaled by functions of the
scattering \cite{Laenen:1998qw},
\bea\label{eq:Na-Nb-1pi}
N_a \ &=&\ N\,\frac{ \beta_b\cdot \beta_r}{\beta_a\cdot \beta_b}\,,
\nn\\[2mm]
N_b \ &=&\ N\, \frac{\beta_a\cdot \beta_r}{\beta_a\cdot \beta_b}\, ,
\eea
where $\beta_{a,b}$ are the velocity four-vectors of the incoming particles,
and $\beta_r$ of the recoil particle in the final state. 

In the same way as in the full partonic calculation, the eikonal cross sections have uncancelled collinear singularities associated with their incoming eikonal lines. As indicated in Eq.~(\ref{eq:tilde-omega}), we regulate collinear singularities  by continuation to $D=4-2\epsilon$ dimensions.  The resulting collinear poles in $\epsilon$ exponentiate, and are cancelled by the collinear poles of the eikonal parton distributions in the denominator of Eq.\ (\ref{eq:tilde-omega}), giving a finite ratio for each $\tilde \omega_{abr}$. We will encounter these cancellations below.   The cancellation follows from standard arguments for factorization \cite{Collins:1985ue,Collins:1988ig,Bodwin:1984hc}, and to leading power in $N$ requires only a simple ratio in moment space, for each choice of incoming eikonal lines. This flavor-diagonal cancellation is a general feature of threshold resummation at leading power in $N$ \cite{Sterman:1986aj,Catani:1989ne,Catani:1990rp}.

Both the eikonal distributions and cross sections are defined as matrix elements of Wilson lines.   For the latter, we use the notation,
\bea
\Phi_\beta^{(R)}\left(\lambda_2,\lambda_1;x\right)
\ \equiv \
{\rm P}\ \exp \left[ -ig\int_{\lambda_1}^{\lambda_2}\ d\lambda\, 
\beta\cdot
A^{(R)}(\lambda\beta+x)\, \right]
\eea
in color representation $R$, where ``P'' denotes path ordering. 
For the annihilation channel,  
$q\bar q \to \beta g $, we use these to construct a three-Wilson line product, which represents the
source of radiation near partonic threshold:
\be
[U_{ {q\bar{q}g} }(x)]_{d;ji}\ \equiv\
{\rm T}\ \left(\;
\left[\, \Phi_{\beta_g}^{(g)}\left(\infty,0;x\right) \, \right]_{d,e}\ u^{(q\bar{q})}_{e,ji}(x)\, \right )\,,
\ee
where ``T'' denotes time ordering, and where
\be
u^{(q\bar{q})}_{e,ji}(x)\  \equiv \  \left [\, \Phi_{\beta_{\bar{q}}}^{(\bar{q})}\left(0,-\infty;x\right)\, \right ]_{jl}
\left(\, T_e^{(q)}\, \right)_{lk}\
\left [\, \Phi_{\beta_q}^{(q)}\left(0,-\infty;x \right)\, \right ]_{ki}\, ,
\ee
with $T_e^{(q)}$ the SU(3) generator in the fundamental representation.
For the Compton channel, $qg \to  \beta q$, the corresponding product is
\bea
[U_{ {qgq} }(x)]_{j;di}\ &=&\
 {\rm T}\ \left(\;
 \left[\, \Phi_{\beta_q}^{(q)}\left(\infty,0;x\right) \, \right]_{jk}\ u^{(gq)}_{k;di}(x)\, \right )\,,
 \nn\\[2mm]
u^{(gq)}_{k,di}(x)\ &=&\  
\left(\, T_e^{(q)}\, \right)_{kl}\ \left [\, \Phi_{\beta_g}^{(g)}\left(0,-\infty;x\right)\, \right ]_{ed}
 \left [\, \Phi_{\beta_q}^{(q)}\left(0,-\infty;x\right)\, \right ]_{li}\,,
\eea
and analogously for the $\bar q g \to \beta \bar q$ channel.
 
In terms of the operators just defined and the transform of Eq.\ (\ref{eq:transform-notation}), the eikonal cross sections are given by
\bea
\tilde \sigma^{\rm (eik)}_{abr} (N/Q, \hat{\eta}, \mu,\epsilon)\ &=&\ \sum_X\ \e^{-N (\beta_r \cdot p_X /Q)}
\left \langle 0 | U^\dagger_{abr}(0)  | X \right \rangle
\left \langle X | U_{abr}(0) | 0 \right \rangle 
\nn\\[2mm]
&=&\ \sum_X
\left \langle 0 | U^\dagger_{abr}(0)  | X \right \rangle
\left \langle X | \e^{-N (i\beta_r \cdot \partial_x /Q)}\  U_{abr}(x) | 0 \right \rangle_{x=0}
\nn\\[2mm]
&=&\ 
\left \langle 0 | U^\dagger_{abr}(0)  \
 \e^{-N (i\beta_r \cdot \partial_x /Q)}\  U_{abr}(x) | 0 \right \rangle_{x=0}\, .
 \label{eq:sig-eik-lines}
\eea
In computing this quantity, an average over 
initial-state colors and a sum over final-state colors (for example, $i,j$ and $d$, respectively, in $[U_{q\bar q g}]_{d;ji}$)  is assumed but suppressed.
The derivative operator in this form is fixed by the velocity of the outgoing Wilson line in the $U_{abr}$ operators.  We will use this feature in the following arguments.

\subsection{Cancellation of final-state interactions}

The outgoing Wilson line effectively changes the exponential of the derivative to the exponential of a covariant derivative,
\bea
{\rm e}^{-{iN\beta_r\cdot \partial_x\over Q}}\,
\Phi_{\beta_r}^{(r)}(\infty,0;x)\
=\
\Phi_{\beta_r}^{(r)}(\infty,0;x)\;
{\rm e}^{-{iN\beta_r\cdot D^{(r)}(A(x))\over Q}}\, ,
\label{across}
\eea
in terms of the covariant derivative $\beta_r\cdot D^{(r)}(A(x))= \beta_r\cdot (\partial_x + ig A^{(r)}(x))$, in the color
representation of the recoil parton.
Substituting this basic relation into the expression in Eq.\ (\ref{eq:sig-eik-lines}), we readily find the cancellation of the unitary 
final-state Wilson lines, giving a form
in which all interactions associated with the recoil line are in an exponentiated operator localized at the origin,
\bea
\tilde \sigma^{\rm (eik)}_{abr} (N/Q, \hat{\eta}, \mu,\epsilon)\ &=&
 \langle 0 | u^{ab}{}^\dagger(0) \
[\, {\rm e}^{-{iN\beta_r\cdot D^{(r)}(A(0))\over Q}}\, u^{ab}(0) ] |\, 0  \rangle
\nn\\[2mm]
&=&
 \langle 0 | u^{ab}{}^\dagger(0) \
\left [\, \left ( 1 \ -\ {iN\beta_r\cdot D^{(r)}(A)\over Q} + \dots \right )\, u^{ab}(0) \right ] |\, 0  \rangle
\nn\\[2mm]
&=&\ \sigma^{({\rm eik},0)}_{abr}\, \left( 1 + {\cal O}(\alpha_s) \right )\, .
\label{eq:u-exp-u}
\eea
In the second equality, we exhibit the expansion in $N/Q$ that we will employ below.  Any truncation to a finite power of $N/Q$ corresponds to a local operator of dimension $N$.  In the third equality, we define the normalization factor $\sigma^{({\rm eik},0)}_{abr}$ that appears in the definition of the factor $H_{abr}$ in Eq.\ (\ref{eq:Habr-def}).   Note that $\sigma^{({\rm eik},0)}_{abr}$ is a pure color factor.

In Eq.\ (\ref{eq:u-exp-u}), the cancellation of IR singularities associated with the final state is manifest at any order in the expansion.     The eikonal cross section is an infinite sum of insertions of the covariant derivative in the $\beta_r$ direction on the incoming lines of 
$u^{ab}$ at the origin.    After an order-by-order sum over final states,
there are no singularities at any point where lines are parallel to the outgoing eikonal $\beta_r$, 
simply because there are no eikonal propagators in the perturbative expansion of the right-hand side of Eq.\ (\ref{eq:u-exp-u}).
Collinear singularities associated with the incoming eikonals remain, of course.
As discussed following Eq.\ (\ref{eq:tilde-omega}) all collinear singularities will be eliminated 
in the ratios that define the eikonal hard functions $\tilde \omega_{abr}^{\mathrm{(eik)}}$ in that equation.   We will see how this occurs below.

Logarithms of the moment variable, $N$ 
in $\tilde \sigma^{({\rm eik})}_{abr}$, organized by threshold resummation,
are generated in the usual fashion by an incomplete cancellation of real and virtual emissions.   This mismatch between real and virtual contributions
requires all orders in the expansion of the exponent in Eq.\ (\ref{eq:u-exp-u}), and we will not base our analysis below on this particular form.   The essential conclusion that we employ from this expression is the absence of singular behavior associated with the outgoing eikonal in an alternative expression
for the eikonal cross section, which is fully equivalent to (\ref{eq:u-exp-u}) point-by-point in momentum space.   This is the web expansion, to which we now turn.

\section{Graphical exponentiation and resummation}
   
We now recall the 
representation of eikonal cross sections as exponentials of web functions, and go on to review some of the relevant properties of web functions, in particular their renormalization-scale independence and absence of subdivergences.   We explain the relation between perturbative resummation and web exponentiation, and compute the exponent at lowest order in the running coupling.   Using the renormalization group independence of webs, we see
already at this order that power corrections induced by the running coupling, or finite gluon mass, are characterized by even powers of $N/p_T$.   In the following section, we go on to extend this result to all orders.

\subsection{Graphical exponentiation and web functions}

The value of the expression for the cross section, Eq.\ (\ref{eq:u-exp-u}) is that it shows that all infrared singularities associated with couplings to the outgoing line cancel in the eikonal cross section.  This result follows because the moment variable is defined by the same velocity vector, $\beta_r$ that defines the single outgoing Wilson line for the eikonal cross sections.
In this section, we will exploit this cancellation, using  the exponentiated web representation of these eikonal cross sections based on three eikonal lines.   
A specific discussion of exponentiation for products of three Wilson lines at the amplitude level can be found in \cite{Berger:2003zh}, and an eikonal cross section written as in Eq.\ (\ref{eq:sig-eik-lines})  or (\ref{eq:u-exp-u}) falls into the general arguments of \cite{Mitov:2010rp} as long as the integral over phase space is symmetric for all final-state partons. This is the case for the cross section defined by Eq.\ (\ref{eq:sig-eik-lines}).  
It may therefore be written in the form
\be
\tilde \sigma_{abr}^{({\rm eik})}(N/Q,\hat\eta,\mu,\epsilon)\ =\ \tilde \sigma^{\rm (eik,0)}_{abr}\, {\rm e}^{E_{abr}(N/Q,\hat\eta,\mu,\epsilon)}\, ,
\label{eq:tilde-sigma-eik}
\ee
where
\bea
E_{abr}(N/Q,\hat\eta,\mu,\epsilon)\ &=&\ 
\int {d^Dk\over (2\pi)^D}\ \left(\, {\rm e}^{-N{\beta_r\cdot k \over  Q}} -
1\, \right) \, \theta \left( \frac{Q}{\sqrt{2}}  - k^+ \right)\, 
\theta\left( \frac{Q}{\sqrt{2}}  - k^- \right)
\nn\\[2mm]
\ & \times &\
w_{abr} \left(\left\{{\beta_i\cdot k\,  \beta_j\cdot k\over
\beta_i\cdot\beta_j}\right\},k^2,\mu^2,
\alpha_s(\mu^2)\right)\, .
\label{eq:sig-abc}
\eea
For this three-eikonal case, the web function $w_{abr}$ is a scalar in group space, and the overall factor $\tilde \sigma^{\rm (eik,0)}_{abr}$ is the zeroth-order eikonal cross section.   
The dependence of the web function on momentum $k$ is fixed by the invariance of straight Wilson lines under rescalings of their defining velocities $\beta_i$.   The explicit Laplace transform corresponds to contributions of all diagrams with at least one parton in the final state, weighted as in Eqs.\ (\ref{eq:transform-notation}) and (\ref{eq:sig-eik-lines}).  The step functions cut off final momenta at the scale of the partonic threshold center-of-mass energy \cite{Laenen:2000ij}.  
Dependence of this cutoff on $Q$
is exponentially suppressed for real radiation.   For brevity of notation, we will suppress these theta functions below.  
The contributions of purely virtual diagrams are summarized by the term with ``1'' that is subtracted in the integrand. The form of the virtual contributions is fixed by imposing that the eikonal cross section vanish at $N=0$, that is, for the fully inclusive eikonal cross section.   

The perturbative expansion of the cross section in Eq.\ (\ref{eq:sig-abc}) is, of course,  collinear-singular, and requires collinear subtractions for its incoming eikonals, as in Eq.\ (\ref{eq:tilde-omega}).  In computing the ratio of eikonal cross sections to eikonal parton distributions in Eq.\ (\ref{eq:tilde-omega}), we use the exponential form for the latter \cite{Laenen:2000ij}, given in $D=4-2\epsilon$ dimensions by
\bea
\tilde{\phi}_{i/i}(N,\mu,\epsilon)\ =\  \exp \left [ \int_0^{\mu^2} \frac{dk^2_T}{k_T^{2(1+\epsilon)}}\, A_i\left(\alpha_s(k_T^2)\right) \, \int_0^1 dz 
\,\frac{ z^N-1}{1-z} \right ]\, ,
\label{eq:eik-phi}
\eea
in terms of the familiar anomalous dimension $A_i(\alpha_s) = C_i\, (\alpha_s/\pi) + \dots$ with $C_q=C_F=4/3$ and $C_g=C_A=3$.   Here, we exhibit the $N$-dependence of the full anomalous dimension 
($-A(\alpha_s)\, \ln (N\e^{\gamma_E})+{\cal O}(1/N)$, with $\gamma_E$ the Euler constant) as an integral over 
a variable $z$.   This form will be convenient in Sec.\ \ref{sec44}.

Substituted into Eq.\ (\ref{eq:tilde-omega}), the eikonal cross section (\ref{eq:tilde-sigma-eik}) and subtractions (\ref{eq:eik-phi}) give for the eikonal hard-scattering cross section,
\be
\tilde \omega_{abr}^{\mathrm{(eik)}}(N,Q,\hat\eta,\mu) \ =\ 
H_{abr}(p_T,\hat{\eta},\mu)\, \sigma^{({\rm eik},0)}_{abr} \ \e^{ \hat E_{abr}(N/Q,\hat\eta,\mu) }\,,\
\ee
where
\bea
\hat E_{abr}(N/Q,\hat\eta,\mu) &=&
\int {d^Dk\over (2\pi)^D}\ \left(\, {\rm e}^{-N{\beta_r\cdot k \over 
Q}} - 1\, \right)\; w_{abr} \left(\left\{{\beta_i\cdot k\,  \beta_j\cdot k\over
\beta_i\cdot\beta_j}\right\},k^2,\mu^2,
\alpha_s(\mu^2)\right)
\nn\\[2mm]
&- &
\sum_{i=a,b} \int_0^{\mu^2} \frac{dk^2_T}{k_T^{2(1+\epsilon)}}\, A_i\left(\alpha_s(k_T^2)\right) \, \int_0^1 dz 
\,\frac{ z^{N_i}-1}{1-z} \, ,
\label{eq:E-10}
\eea
In this form, the cancellation of collinear singularities appears in the exponent, the logarithm of the cross section, which as we have seen is given at each order by
a sum of web diagrams.  The integrals of the web diagrams will thus automatically generate a single collinear-singular integral, exactly matching the collinear
integrals of the eikonal parton distributions, specified by the anomalous dimensions $A_a$.  We will organize these integrals below in a manner that exhibits collinear-finite power corrections in $N/Q$,
but the cancellation of leading-power collinear singularities is guaranteed by factorization, as discussed below Eq.\ (\ref{eq:tilde-omega}).

 Leading powers in $Q$ come from the range of $k$ where $N\beta_r\cdot k/Q > 1$ and real-gluon emission is exponentially suppressed.    Potential contributions  at nonleading powers in $N/Q$ arise from the region  $N\beta_r\cdot k/Q <1$, and can be isolated by expanding the exponential,  as in Eq.\ (\ref{eq:u-exp-u}), for the full eikonal cross section. We will analyze this expansion  in Sec.\ \ref{sec:IR-PC},
 after discussing relevant properties of the web functions.

\subsection{Webs, their renormalization and absence of subdivergences}
\label{sec:no-subs}

The web functions in Eq.\ (\ref{eq:sig-abc}) can be defined recursively, in perturbation theory, with a general form given by  \cite{Mitov:2010rp},
\bea
w^{(N+1)} = 
\sum_{D^{(N+1)}}  D^{(N+1)} - \left[\, \sum_{m=2}^{N+1} \frac{1}{m!}\,\,
\sum_{i_m=1}^N \dots \sum_{i_1=1}^N\,
w^{(i_m)}\, w^{(i_{m-1})} \dots w^{(i_1)} \right]^{(N+1)}\, ,
\label{eq:first_wN+1}
\eea
where the $D^{(N+1)}$ make up the full set of eikonal diagrams at the $(N+1)$st order in $\alpha_s$, and the web function $w^{(M)}$ is the sum of web diagrams at $M$th order.  The square bracket with superscript $(N+1)$ indicates that the $(N+1)$st order only is kept in the multiple sum over 
lower-order webs.  At first order, the webs are given by the sum of single-gluon exchanges between Wilson lines, and Eq.\ (\ref{eq:first_wN+1}) determines all higher orders. 
We have suppressed momentum dependence, integrations and
parton labels, but we emphasize that this form applies to any eikonal cross section with symmetric phase space, not simply to the three-eikonal case at hand.  In the general case, with four or more eikonals, the web functions are group space matrices.  

The webs for electroweak boson production cross sections in our case are given by diagrams with 
three-eikonal irreducibility, which differ from their standard forms by the subtraction of certain color factors 
\cite{Gardi:1999dq,Berger:2003zh}
(maximally nonabelian configurations are left).  Their propagators are the usual causal free Green functions for quarks and gluons, and the usual eikonal denominators for Wilson lines.   Because the propagators are standard,  web diagram 
integrals over the components opposite to $\beta_r$ can be carried out in light-cone perturbation theory, 
and the cancellation of final states holds diagram-by-diagram in the inclusive cross section \cite{Collins:1988ig}.  The scale invariance of eikonal diagrams implies
the homogeneous dependence 
on the variables involving the Wilson line velocities, $\beta_i^\mu$ shown in Eq.~(\ref{eq:sig-abc}). 

The integral of the virtual web function over momentum $k$ in Eq.\ (\ref{eq:E-10}) requires ultraviolet renormalization, which is additive, and corresponds to the renormalization of the composite three-eikonal vertex that defines the cross section.   At fixed momentum $k$, however, the web functions are unrenormalized, that is, renormalization scale invariant,
\bea
\mu \frac{d}{d\mu} 
w_{abr}
\left(\left\{{\beta_i\cdot k\,  \beta_j\cdot k\over
\beta_i\cdot\beta_j}\right\},k^2,\mu^2,
\alpha_s(\mu^2)\right)\ =\ 0\, .
\label{eq:w-rg}
\eea
This is a consequence of the lack of an analog of wave function renormalization for Wilson lines \cite{Dotsenko:1979wb}.  Equation (\ref{eq:w-rg}) holds for both massive and massless Wilson lines.

In addition to their renormalization properties, webs also have the important property that they are free of subdivergences, aside from logarithms that can be absorbed into the running coupling through Eq.\ (\ref{eq:w-rg}).  This means that IR divergences in the integral over $k$ arise only when both $k^2$ and one of the combinations 
$\beta_i\cdot k\, \beta_j\cdot k$ vanish.   For the cross section at hand, these are soft and collinear limits of web diagrams. 

To make the discussion specific, let us 
consider a subspace $S_L$ of the loop momenta of web $w^{(N)}$ in an $L$th-order subdiagram, $L<N$.
We denote by ``$S_L\to 0$'' a configuration where all of the loop momenta of the subdiagram are soft or collinear.   We want to show that
\bea
w^{(N)} \ \stackrel{S_L\to 0}{\longrightarrow }\ 0\, , 
\label{eq:W-single}
\eea
where the zero on the right-hand side refers to the absence of an infrared subdivergence.   The lack of subdivergences, Eq.\ (\ref{eq:W-single}), certainly holds at $N=1$, which is the case of a single gluon. We next assume that it applies for webs up to order $N$.

Extending  Eq.\ (\ref{eq:W-single}) to order $N+1$ is readily seen as consequence of the order-by-order factorization properties of cross sections in such limits.  
We consider the $L$-loop on-shell, soft or collinear limit of $w^{(N+1)}$ as given by Eq.\ (\ref{eq:first_wN+1}).  For the first term on the right-hand side of (\ref{eq:first_wN+1}), general factorization properties give,
\bea
\sum_{D^{(N+1)}}  D^{(N+1)} \ \stackrel{S_L\to 0}{\longrightarrow }\  \sum_{D^{(L)}}  D^{(L)} \ \sum_{D^{(N+1-L)}}  D^{(N+1-L)} \, ,
\label{eq:D-fact}
\eea
where corrections are again nonsingular.  
To treat the second term, consisting of sums of lower-order webs, we use our assumption that up  to order $N$, web functions only diverge when the entire web becomes collinear or soft
altogether.    Thus, the limit $S_L\to 0$ can only be realized when a single web in order $L$ has loop momenta all in $S_L$, or when a set of web functions whose union is a diagram in the sum over $D^{(L)}$ have loop momenta that together are in $S_L$.   
Factoring these webs algebraically from the sum in (\ref{eq:first_wN+1}), we find
\bea
w^{(N+1)} \
&\stackrel{S_L\to 0}{\longrightarrow }& \ 
 \sum_{D^{(L)}}  D^{(L)}  \sum_{D^{(N+1-L)}}  D^{(N+1-L)}
 \nn\\[2mm]
&\ &\hspace*{-1cm} \ -  
 \ \left \{  w^{(L)}   + \left[ \sum_{m'=2}^{L} \frac{1}{m'!}\,\,
\sum_{j_{m'}=1}^L \dots \sum_{j_1=1}^L\,
w^{(j_{m'})}\, w^{(j_{m'-1})} \dots w^{(j_1)} \right]^{(L)} \right \}
 \times \sum_{D^{(N+1-L)}} D^{(N+1-L)}
\nn\\[2mm]
&=& 0\, ,
\eea
where we have also used Eq.\ (\ref{eq:first_wN+1}) to identify the full set of $(N+1-L)$th order diagrams as a sum of webs.  
Referring above to Eq.\ (\ref{eq:first_wN+1}),  the definition of the web function at $(N+1)$st order,
we see that the term in curly brackets cancels the sum over the $D^{(L)}$, by the definition of the web function $w^{(L)}$ for $L\le N$.   The subdivergences of $w^{(N+1)}$ thus also cancel in the sum over webs, which is what we set out to show.

\subsection{Perturbative resummations and eikonal power corrections}
 
Before studying power corrections, we return to the full form of the eikonal cross section in Eq.\ (\ref{eq:sig-abc}) and 
the hard-scattering function (\ref{eq:E-10}), and remark that perturbative and nonperturbative contributions separate naturally in the integral over $k$.   Perturbative resummation of logarithms in $N$ can be derived from the range of momenta $k$ for which $\beta_r\cdot k \ge Q/N$.   In this region, the term due to real emission is exponentially suppressed, and logarithms in $N$ arise entirely from the contributions of virtual diagrams, given by the function $w_{abr}$, with subtractions specified as in Eq.\ (\ref{eq:E-10}) in the range $1-z>1/N$, where $z^N\ll 1$.   The results for direct photons, developed from this point of view, are found in Ref.\ \cite{Laenen:2000ij}.   Alternative (and more familiar) approaches to resummation for single-particle production 
are based on refactorizations of the hard scattering function, leading to evolution 
equations \cite{Catani:1999hs,Kidonakis:1999hq,Kidonakis:1999ur,Gonsalves:2005ng,deFlorian:2005fzc,Catani:2013vaa,Hinderer:2018nkb}, and on related analyses in Soft Collinear Effective Theory \cite{Becher:2009th}.  These analyses go beyond the eikonal approximation that we are discussing here.   From a practical point of view, a purely eikonal analysis of resummation would miss contributions associated with partonic spin and kinematics starting at next-to-leading logarithms.   For the reasons discussed at the beginning of Sec.\ \ref{sec:eikonal}, 
however, we believe that the eikonal approximation captures the leading nonperturbative behavior that appears in powers of $N/Q$.    

In what follows, we take this viewpoint, and concentrate on the range of $k$ where $\beta_r\cdot k \le Q/N$, and hence for which an expansion of the exponential $\exp [-N\beta_r\cdot k/Q]$ converges rapidly
in Eqs.\ (\ref{eq:sig-abc}) and (\ref{eq:E-10}).    In this range, which is not normally included in perturbative resummation,  there is an interplay between real and virtual corrections at low momentum scales.  To study the integral in this region, the properties of the web functions derived above will be useful.

\subsection{The lowest order web \label{sec44}}

We illustrate these considerations with the lowest-order web function for $q\bar q\to \beta+g$, given by the interference terms between gluon emission from each of the three Wilson lines.
The $C_F$ part of the calculation proceeds in much the same way as the one for the inclusive Drell-Yan cross section 
in Ref.~\cite{Laenen:2000hs}. 
It is natural to specialize to the frame where $\beta_a^\mu$ and $\beta_b^\mu$ are incoming along the 3 axis.  We continue to normalize velocities as $\beta_{a,b}^0=1$, so that
\bea
\beta_a^\mu\ &=&\ \sqrt{2}\, \delta_{\mu +}\, ,
\nn\\[2mm]
\beta_b^\mu\ &=&\ \sqrt{2}\, \delta_{\mu -}\, ,
\label{eq:beta-ab}
\eea
giving $\beta_a\cdot \beta_b=2$ and
\bea
{\beta_a\cdot k\,  \beta_b\cdot k\over
\beta_a\cdot\beta_b}\ =\
\frac{k^2+k_T^2}{2}\, .
\label{eq:k2kt2}
\eea
It is now convenient to project the recoil velocity in the following way:
\bea
\beta_r^\mu\ =\ \frac{\beta_r\cdot \beta_b}{\beta_a\cdot\beta_b}\ \beta_a^\mu\ +\ \frac{\beta_r\cdot \beta_a}{\beta_a\cdot\beta_b}\ \beta_b^\mu\ + \beta_{r,T}\, .
\label{eq:beta-r-expand}
\eea
A bit of algebra then gives for the first-order web function in the $q\bar{q}$ channel:
\bea
w_{q\bar q g}^{(1)}\ &=&\ 4\pi g^2\,\mu^{2\epsilon}\,\delta(k^2)\ \left ( C_F \frac{\beta_a\cdot \beta_b}{\beta_a\cdot k \beta_b\cdot k}\ +\ \frac{C_A}{2}\, \frac{\beta_a\cdot \beta_b}{\beta_a\cdot k \beta_b\cdot k}\,
\frac{\beta_{r,T}\cdot k_T}{\beta_r\cdot k } \right )
\nn\\[2mm]
&\equiv&\ C_F\, u_{q\bar q g}^{(1)}\ +\ C_A\, v^{(1)}_{q\bar q g}\, .
\label{eq:lo-web-qqbarg}
\eea
 The term $u^{(1)}_{q\bar q g}$, multiplied by $C_F$ contains all collinear singularities in the web contribution to Eq.\ (\ref{eq:E-10}). It is easy to see that with $A(\alpha_s) = C_F \alpha_s/\pi + \dots$, these singularities are cancelled by the collinear subtraction, as they must be.  The term $w_{q\bar qg}^{(1)}$, proportional to $C_A$ is by itself collinear singular when $k$ is collinear to $\beta_r$, but this singularity cancels between real and virtual corrections in Eq.\ (\ref{eq:E-10}).   We may thus substitute the web function $w^{(1)}_{q\bar q g}$ of Eq.\ (\ref{eq:lo-web-qqbarg}) in Eq.\ (\ref{eq:E-10}) and evaluate the resulting integral with $D=4$.   In the large-$N$ limit, we can approximate the $z$ integrals in the subtraction term of (\ref{eq:E-10}) 
 by $\ln N_i {\mathrm{e}}^{\gamma_E}\equiv \ln \bar N_i$.   
 We then have, at lowest order in $\alpha_s$,
 \bea
\hat E^{(1)}_{abr}(N/Q,\hat\eta,\mu)\ &=&\ C_F \int {d^4 k\over (2\pi)^4}\ \left(\, {\rm e}^{-N{\beta_r\cdot k \over Q}} - 1\, \right) u^{(1)}_{abr}(k)
\nn\\[2mm]
&+ &
\int_0^{\mu^2} \frac{dk^2_T}{k_T^{2}}\, C_F\, \frac{\alpha_s(k_T^2)}{\pi}  \, \ln \left ( \frac{\bar N\beta_{r,T}}{2} \right ) 
\nn\\[2mm]
&+&\ C_A \, \int {d^4 k\over (2\pi)^4}\ \left(\, {\rm e}^{-N{\beta_r\cdot k \over Q}} - 1\, \right) v^{(1)}_{abr}(k)
\nn\\[2mm]
&\equiv&\ C_F\, U^{(1)}_{q \bar q g}(N/Q,\hat\eta,\mu)\ +\ C_A\, V^{(1)}_{q \bar q g}(N/Q,\hat\eta,\mu)\, ,
\label{eq:E-u-v}
\eea
 where we have used the choices of moment variables for single-particle cross sections given in Eq.\ (\ref{eq:Na-Nb-1pi})
  and that, in the normalization of Eq.\ (\ref{eq:beta-ab}), $\beta_{r,T}^2 = \beta_b\cdot\beta_r\, \beta_a\cdot \beta_r$.

It is straightforward to reduce the resulting expression for the $C_F$ term in Eq.\ (\ref{eq:E-u-v}) to a single integral over $k_T$, in the limit that $N\gg 1$ for values of $k_T$ small enough that $Nk_T/Q \le 1$.  
We begin by expressing
the mass-shell delta function in the web as $(1/2k^+)\delta(k^- - k_T^2/2k^+)$ to do the $k^-$ integral, 
imposing the limits on the $k^+$ and $k^-$ integrals shown
explicitly in Eq.\ (\ref{eq:sig-abc}).  
In the large-$N$ limit, we can  then perform the $k^+$ integral to get a Bessel function, up to terms that are exponentially suppressed in $N$, using the relation
\bea
\int_{k_T^2/\sqrt{2}Q}^{Q/\sqrt{2}} \frac{dk^+}{2k^+}\, \left ( {\mathrm{e}}^{-\frac{N}{Q} \left( \beta_r^- k^+ + \beta_r^+ \frac{k_T^2}{2k^+} \right)} -1 \right )
\ = \
K_0 \left( \frac{ N\beta_{r,T}k_T}{Q} \right )\ + \ \ln\left(\frac{k_T}{Q}\right)\ +\ {\cal O} \left ( {\mathrm{e}}^{-N} \right )\,.
 \label{eq-K0-approx}
 \eea
 We note that this approximation holds for any $k_T$ less than $Q$, and that, in fact, 
 the function $K_0(z)$ behaves as $z^{-1/2}e^{-z}$ for large $z$.
 For the function $U^{(1)}_{q \bar q g}(N/Q,\hat\eta,\mu)$, we then find
  \bea
 U^{(1)}_{q \bar q g}(N/Q,\hat\eta,\mu)\ =\ 2 \int_0^{Q^2} \frac{dk_T^2}{k_T^2}\,  \frac{\alpha_s(k_T^2)}{\pi} \,
 \left [ I_0 \left ( \frac{ N\beta_{r,T} k_T}{Q} \right)\, K_0 \left ( \frac{ N\beta_{r,T} k_T}{Q} \right)\ +\ \ln \left ( \frac{\bar N\beta_{r,T}k_T}{2Q} \right ) \right ] \, ,
 \nn\\
  \label{eq:U1-result}
  \eea
  where $I_0(x)$ is a standard Bessel function, for which $I_0(0)=1$.   In the range for which $k_T$ is small enough that $Nk_T/Q<1$, standard
  expansion formulas for the functions  
  $K_0$ and $I_0$ show that power corrections involve only even powers of $N/Q$.   As noted above, for large $N$ and
  $k_T$, the Bessel function $K_0$ decreases exponentially.   In Eq.\ (\ref{eq:U1-result}), this leaves the exponentiating double-logarithmic corrections of threshold resummation.
    
  For the second, $C_A$, term in Eq.\ (\ref{eq:E-u-v}),  it is easier to compute the derivative with respect to $N$, which cancels the $\beta_r\cdot k$ denominator in
  Eq.\ (\ref{eq:lo-web-qqbarg}).   We find that
  \bea\label{Vterm}
  \frac{d}{dN}\, V^{(1)}_{q \bar q g}(N/Q,\hat\eta,\mu)\  =\ \frac{2\beta_{r,T} }{Q}
  \int_0^{Q^2} \frac{dk_T^2}{k_T}\, \frac{\alpha_s(k_T^2)}{\pi} \,
  I_1 \left ( \frac{ N\beta_{r,T} k_T}{Q} \right)\, K_0 \left ( \frac{ N\beta_{r,T} k_T}{Q} \right)\,.
\eea
Although $N$ is large, in the region of interest we can still expand the Bessel functions in terms of their arguments, leading to an expression that begins at power $N/Q$ times even powers of $N/Q$.   
Such an expansion is legitimate since the integrand in Eq.~(\ref{eq:E-u-v}) vanishes at $N=0$ for any $k_T$.
This also means that there is no term constant in $N$ in this expansion. 
Assuming $V^{(1)}$ has an expansion in the same variables, it must 
start at $N^2/Q^2$, just as for $U^{(1)}$.   In summary,
the lowest-order behavior in $N/Q$ for the functions $U^{(1)}$ and $V^{(1)}$ in the region $N\gg1$, $Nk_T/Q \le 1$ is, 
\bea
U^{(1)}_{q \bar q g}(N/Q,\hat\eta,\mu)\ &=&\  \frac{1}{2}
\left (  \frac{ N\beta_{r,T}}{Q} \right )^2\, \int_0 dk_T^2\,  \frac{\alpha_s(k_T^2)}{\pi} \,\left[ 1-
2 \ln \left ( \frac{\bar N\beta_{r,T}k_T}{2Q} \right ) \right] \ +\ {\cal O}(N^4/Q^4)\, ,
\nn\\[2mm]
V^{(1)}_{q \bar q g}(N/Q,\hat\eta,\mu)\ &=&\  -\, \frac{1}{2} \left (  \frac{ N\beta_{r,T}}{Q} \right )^2\,  \int_0 dk_T^2\,
 \frac{\alpha_s(k_T^2)}{\pi} 
\ln \left (  \frac{ \bar N\beta_{r,T}k_T }{Q} \right )\ +\ {\cal O}(N^4/Q^4)\, .
\label{eq:U1-V1-expand}
\eea
Here we see explicitly the absence of linear powers.  
The part with $U^{(1)}_{q \bar q g}$ is very similar to the result for inclusive Drell-Yan obtained in~\cite{Laenen:2000hs}. 
As noted, the expansions of the relevant Bessel functions generate even powers to all orders in this expansion. Although we are not able of course to do the $k_T$ integrals in~(\ref{eq:U1-result}) and~(\ref{Vterm}) in closed
form, this feature is completely general. 
We have checked that it equally applies in the case of $qg$ scattering, despite the fact
that the collinear subtraction here contains two separate terms, one with $C_F$ and one with $C_A$. 

We have used $k_T$ as the scale of the running coupling in all of these expressions.   This is a consistent choice, given the scale-invariance of web functions, Eq.\ (\ref{eq:w-rg}) and the
lack of subdivergences in webs discussed above.  The ambiguity present in resummed perturbation theory is then evident in Eq.\ (\ref{eq:U1-V1-expand}), due to the Landau singularity in  $\alpha_s(k_T^2)$.   These singularities may be organized, for example, in Borel form \cite{Korchemsky:1994is}, by using the representation 
\bea
\alpha_s( k_T^2)\ =\ \int_0^\infty d\sigma\, \left(\frac{k_T^2}{\Lambda^2}\right)^{-\sigma\beta_0/4\pi}\,  ,
\eea
in Eq.\ (\ref{eq:U1-V1-expand}),
where $\beta_0=11-2N_f/3$ (with $N_f$ the number of flavors) is the lowest-order coefficient of the QCD $\beta$ function. 
In any case, the absence of a linear $N/Q$ correction is the main result, familiar from Ref.\ \cite{Beneke:1995pq}.  We extend this result to all orders
in the following section.

In closing this section,
we can check that  analogous exponentiating power corrections appear at lowest order and fixed coupling if we provide the gluon with a mass, as in Ref.\ \cite{Caola:2021kzt}.    This can be done by making the simple replacement
\bea
\delta(k^2)\ \rightarrow\ \delta(k^2-\lambda^2)
\eea
in the lowest-order web, Eq.\ (\ref{eq:lo-web-qqbarg}), also modifying $k_T^2$ to $k_T^2+\lambda^2$ in the subtraction term.
The resulting corrections also involve only even powers of $\lambda$.

\section{Infrared momenta and power corrections}
\label{sec:IR-PC}

It is our goal in this section to demonstrate that the lack of linear powers in $N/Q$ extends to all orders in the web expansion.
The web expansion includes information on power corrections, derived 
by expanding the exponent in the region $\beta_r \cdot k \le Q/N$ in Eq.\ (\ref{eq:sig-abc}) for the eikonal cross section.   We have seen in Eq.\ (\ref{eq:u-exp-u}) that in the full moment integral all final-state interactions generated by the outgoing Wilson line $\Phi_{\beta_r}^{(r)}$ can be replaced by an expansion in terms of operators that are local at the three-eikonal vertex (the ``origin").  These operators are generated from the exponential of the covariant derivative 
$\beta_r\cdot D^{(r)}(A(0))$ multiplied by $N/Q$.   
Hence, each inverse power of $Q$ in the expansion comes with a polynomial of the same order in the vector $\beta_r^\mu$.   

We will begin our general
study of power corrections with contributions that are linear in the vector $\beta^\mu_r$. 
We will see that despite its overall factor of $N/Q$ the $N\beta_r\cdot k/Q$ term is actually leading power and collinear singular.  These singular terms, however, are cancelled in 
the eikonal hard-scattering function $\tilde\omega_{abr}^{\mathrm{(eik)}}$, leaving a residue that is free of linear, $1/Q$ corrections, and which begins at order $N^2/Q^2$.   We will then go on to show that this is the case for any power in the expansion in $N\beta_r\cdot k/Q$.
In the following arguments, we use our ability to identify possible sources of logarithmic singularities with well-characterized regions in web loop momenta.  We will find power corrections beginning at $N^2/Q^2$
 associated with the low-scale behavior of the QCD running coupling
at all orders.

\subsection{Web functions in the power expansion}

Perturbative resummations that are sensitive to the Landau singularity at power $1/Q$ must come from momentum integrals that are pinched at on-shell configurations in Eq.\ (\ref{eq:sig-abc}).   The web function itself is finite order-by-order at fixed nonvanishing values of its arguments.   It is only when a subset of the invariants vanish that there is the possibility of generating logarithms of these invariants.  

We begin our discussion of power corrections with those associated with the first power in the expansion of the the exponential $E_{abr}$ in Eq.\ (\ref{eq:sig-abc}) in $N\beta_r\cdot k/Q$, which we consider to all orders in the strong coupling.  
In view of our observation that dependence on the outgoing velocity $\beta_r^\mu$ is local at the three-eikonal vertex in our cross section to any fixed power in $1/Q$, all true final-state interactions, associated with the outgoing Wilson line, cancel in the web function just as in the full moment of the eikonal cross section.    We denote this result as follows,
\bea
E^{[1]}_{abr}(N/Q,\hat\eta,\mu,\epsilon)\ &\equiv&\ -\ N
\int {d^Dk\over (2\pi)^D}\ \left(\, \frac{\beta_r\cdot k}{Q}\, \right)
w_{abr}
\left(\left\{{\beta_i\cdot k\,  \beta_j\cdot k\over
\beta_i\cdot\beta_j}\right\},k^2,\mu^2,
\alpha_s(\mu^2)\right)
\nn\\[2mm]
&\equiv&\ -\ N
\int {d^Dk\over (2\pi)^D}\ \left(\, \frac{\beta_r\cdot k}{Q}\, \right)
\rho_{abr}^{[1]}
\left(\left\{{\beta_i\cdot k\,  \beta_j\cdot k\over
\beta_i\cdot\beta_j}\right\},k^2,\mu^2,
\alpha_s(\mu^2)\right)\, , \;\;\;
\label{eq:E-1}
\eea
where in the second form, the function  
$\rho_{abr}^{[1]}$
represents the sum of contributions to the full web function $w_{abr}$ 
 that remain after the sum over final states, that is, when integrated over final-state momentum $k$, weighted by $\beta_r\cdot k$.   The superscript $[1]$ in square brackets for the web integrand and the exponent refers to first order in the power expansion, and remains to all orders in the coupling expansion.   We note that at this point, we are working with the exponent of the full cross section rather than the subtracted exponent $\hat E_{abr}$
of Eq.\ (\ref{eq:E-10}).
   We will return to the role of the initial-state collinear subtractions that relate the two below.
 
  Given the expression for the cross section in (\ref{eq:sig-abc}), in which final-state interactions have cancelled, we can conclude that the function  
 $\rho_{abr}^{[1]}$
 in Eq.\ (\ref{eq:E-1}) is much simpler than the full web function.  Indeed, because $E^{[1]}_{abr}$ can depend on $\beta_r$ only through a single linear factor, we can write the residual web function  
 $\rho_{abr}^{[1]}$
 in the form
\bea
\beta_r\cdot k\, 
\rho_{abr}^{[1]}
\left(\left\{{\beta_i\cdot k\,  \beta_j\cdot k\over
\beta_i\cdot\beta_j}\right\},k^2,\mu^2,
\alpha_s(\mu^2)\right)
&=&
\beta_r\cdot k\ 
g_{ab} \left( {\beta_a\cdot k\,  \beta_b\cdot k\over
\beta_a\cdot\beta_b},k^2,\mu^2,
\alpha_s(\mu^2)\right)
\nonumber\\[2mm]
&+&{\beta_r\cdot \beta_a\, \beta_b\cdot k \over \beta_a\cdot \beta_b} \
g_{ar}
\left( {\beta_a\cdot k\,  \beta_b\cdot k\over
\beta_a\cdot\beta_b},k^2,\mu^2,
\alpha_s(\mu^2)\right)
\nn\\[2mm]
&+& {\beta_r\cdot \beta_b\, \beta_a\cdot k \over \beta_a\cdot \beta_b}\
g_{br}
\left( {\beta_a\cdot k\,  \beta_b\cdot k\over
\beta_a\cdot\beta_b},k^2,\mu^2,
\alpha_s(\mu^2)\right)
\, ,
\nn\\
\label{eq:linear-beta-c}
\eea
where the functions $g_{ij}$ have the same mass dimension as the webs, $-D$ in $D$ dimensions.  
Note that the renormalization group invariance of the full web, Eq.\ (\ref{eq:w-rg}), is inherited by these functions individually.   This is because the local subtractions involved in renormalization are independent of the expansion in $N\beta_r\cdot k/Q$.  We therefore have,
\bea
\mu \frac{d}{d\mu} g_{ij}\ =\  0\, ,
\label{eq:g-rg}
\eea
which we will use below.

\subsection{Web structure at lowest order and beyond} 
\label{sec:lo-beyond}

Before turning to the identification of power corrections, we would like to 
see how the decomposition of the web functions  
$\rho_{abr}^{[1]}$
in Eq.\ (\ref{eq:linear-beta-c}) happens in perturbation theory.   
To this end, we return to the lowest order addressed in Sec.~\ref{sec44}, where the web is a single gluon, exchanged 
among the three Wilson lines. This time, we only keep the linear term in the expansion in $N\beta_r\cdot k/Q$. 

To illustrate the lowest order, we consider $q+\bar q \to \beta+g$, represented here, of course, by eikonal lines in the appropriate color representations.   There are six terms (two pairs of three) corresponding to the interference between gluon emission from pairs of different webs.
Factoring out the zeroth-order color factor ($C_F N_c$), which is not part of the web, we find, in $D=4-2\epsilon$ dimensions,
\bea
\beta_r\cdot k\,
\rho_{q\bar q g}^{[1],(1)}
&=&\
2\, (2\pi)g^2\mu^{2\epsilon}\, \, \delta_+(k^2)\ \beta_r\cdot k 
\nn\\[2mm]
&\ &\  \times\ \left \{ \left [ C_F- \frac{C_A}{2} \right ]\, \frac{ \beta_a\cdot \beta_b}{\beta_a\cdot k \beta_b\cdot k} \,
+
\frac{C_A}{2}\, \left [{\beta_r\cdot \beta_a\,  \over \beta_a\cdot k \beta_r\cdot k}
+ {\beta_r\cdot \beta_b\,  \over \beta_b\cdot k \beta_r\cdot k}\,
\right ]\,  \right \}
\nonumber\\
\nn\\[2mm]
&=&\ 2\,
(2\pi)g^2\mu^{2\epsilon}\, \, \delta_+(k^2)\ \frac{ \beta_a\cdot \beta_b}{\beta_a\cdot k \beta_b\cdot k} 
\nn\\[2mm]
&\ &\ \times\ \left \{ \beta_r\cdot k\, \left [ C_F- \frac{C_A}{2} \right ]\, 
+
\frac{C_A}{2}\, \left({\beta_r\cdot \beta_a\, \beta_b\cdot k \over \beta_a\cdot \beta_b}
+ {\beta_r\cdot \beta_b\, \beta_a\cdot k \over \beta_a\cdot \beta_b}\,
\right )\,  \right \}\, ,
\label{eq:linear-beta-c-qqg-lo}
\eea
where the superscript with square brackets, $[1]$, refers to the power expansion in $N/Q$, and the parentheses 
in superscript $(1)$ refer to the order in $\alpha_s$. In the second equality in Eq.\ (\ref{eq:linear-beta-c-qqg-lo}),
we match the kinematic factors involving $\beta_r$ to the form in Eq.\ (\ref{eq:linear-beta-c}).  
This enables us to read off the functions $g_{ij}$ at lowest order.
In this case, two functions are equal, 
\bea
g^{(1)}_{q\bar q} &=&  2\, \left [ C_F- \frac{C_A}{2} \right ]\, (2\pi)g^2\mu^{2\epsilon}\, \delta_+(k^2)\ \frac{ \beta_a\cdot \beta_b}{\beta_a\cdot k \beta_b\cdot k}\, ,
\nn\\[2mm]
g^{(1)}_{q g}\ &=&\ g^{(1)}_{\bar q g}\ =\  2\, \frac{C_A}{2}\, (2\pi)g^2\mu^{2\epsilon}\, \delta_+(k^2)\ \frac{ \beta_a\cdot \beta_b}{\beta_a\cdot k \beta_b\cdot k}\, .
\label{eq:g-h-qqbar-g}
\eea
All of these functions show collinear singularities at $k_T=0$, with $k_T$ defined relative to the scattering axis of the initial state.   After the cancellation of final states, however, collinear singularities at order $\alpha_s$ occur only with color factor $C_F$, corresponding to the incoming eikonals in the quark pair representations, which are present in $g_{q\bar q}^{(1)}$.   The $C_A$ terms must be free of collinear singularities, which in fact they are, since we can readily check that the terms proportional to $C_A$ in (\ref{eq:linear-beta-c-qqg-lo}) cancel when $k$ is in the $a$ or $b$ direction.  
These observations are consistent with what we found in Sec.~\ref{sec44}; see Eq.~(\ref{eq:E-u-v}). 

For completeness, we present the corresponding results for the quark-gluon channel,
\bea
g^{(1)}_{q q} &=&  2\, \left [ C_F- \frac{C_A}{2} \right ]\, (2\pi)g^2\mu^{2\epsilon}\, \delta_+(k^2)\ \frac{ \beta_a\cdot \beta_b}{\beta_a\cdot k \beta_b\cdot k}\, ,
\nn\\[2mm]
g^{(1)}_{g q}\ &=&\ g^{(1)}_{q g}\ =\  2\, \frac{C_A}{2}\, (2\pi)g^2\mu^{2\epsilon}\, \delta_+(k^2)\ \frac{ \beta_a\cdot \beta_b}{\beta_a\cdot k \beta_b\cdot k}\, .
\label{eq:g-h-qg-q}
\eea
Here, we readily verify that collinear singularities again appear in the $a$ and $b$ directions with color factors consistent with collinear factorization.

We now show how the same structure is preserved in the function $E_{abr}^{[1]}$ to arbitrary orders in the coupling.
For any final state, the moment function can be thought of as a sum of two contributions,
\bea
\beta_r\cdot k\ =\ \beta_r\cdot k_{ab} \ +\ \beta_r\cdot k_r\, ,
\label{eq:k-to-kab-kr}
\eea
where $k_{ab}$ is the total momentum flowing out of all vertices on the incoming eikonals, and 
$k_r$ is the total momentum flowing out of the composite vertex at which the outgoing eikonal couples to the incoming eikonals.   
The sum over final states can be taken at fixed $k_{ab}$, and the resulting sum vanishes after integration.
The $\beta_r\cdot k_r$ term on the right-hand side of Eq.\ (\ref{eq:k-to-kab-kr})
 is the inverse of the first outgoing eikonal propagator, corresponding to the gauge field term in the covariant derivative in the expansion of Eq.\ (\ref{eq:u-exp-u}).   This new vertex, with field 
$\beta_r\cdot A^{(r)}$ associated with the three-eikonal vertex, appears for every remaining final state, for which the sum over final states again cancels.

The cancellation can be made manifest by integrating over loop and phase space momentum components projected by $\beta_r$, including  $\beta_r\cdot k$, in light-cone ordered perturbation theory.   This cancellation corresponds to the statement that inserting a complete set of states between the two local operators of Eq.\ (\ref{eq:u-exp-u}), both of which are fixed at the origin, does not change the value of the matrix element.   
The exact cancellation assumes that we integrate over all values of the final-state 
quantity $\beta_r\cdot k$.   In the expressions we consider below, this integration will only be carried out to the order of $Q/N$.   Induced corrections will be suppressed by additional powers of $N/Q$, which we consider to be small.
In this approximation, all final state interactions cancel in the resulting sum over final states, and the result is an expression in which the velocity $\beta_r$ appears only in products with $k$ and/or the incoming velocities, times functions that are boost invariant on the $\beta_a$-$\beta_b$ axis and 
invariant under scalings of $\beta_a,\beta_b$, just as in Eq.\ (\ref{eq:linear-beta-c}).

\subsection{Organizing collinear singularities}

We adopt the frame introduced in Sec.~\ref{sec44}, specifically Eqs.~(\ref{eq:beta-ab}) and~(\ref{eq:k2kt2}). 
In the following, we will relabel the arguments of the web functions as 
$g_{ab}(k_T^2,k^2,\mu^2,\alpha_s(\mu^2))$, and so forth, or for succinctness, simply   $g_{ab}(k_T^2,k^2)$.
With these choices, the first term in the power expansion, using (\ref{eq:linear-beta-c}) for  
$\rho_{abr}^{[1]}$
in Eq.\ (\ref{eq:sig-abc}) for the exponent, is given by
\bea
E^{[1]}_{abr}(N/Q,\hat\eta,\mu,\epsilon)\ &=&\ -\frac{N}{Q}
\int {d^Dk\over (2\pi)^D}\ 
\Bigg [
\beta_r\cdot k\ 
g_{ab} \left( k_T^2,k^2 \right)
+
{\beta_r\cdot \beta_a\, \beta_b\cdot k \over \beta_a\cdot \beta_b} \
g_{ar} \left( k_T^2,k^2 \right)
\nn\\[2mm]
&\ & \hspace{26mm}
+\, {\beta_r\cdot \beta_b\, \beta_a\cdot k \over \beta_a\cdot \beta_b}\
g_{br} \left( k_T^2,k^2 \right)\Bigg ] \, .
\label{eq:E-linear-gs}
\eea
 Because we are at the level of the cross section, the $g_{ij}$ are  collinear divergent when $k$ is parallel to either $\beta_a$ or $\beta_b$.  Because they appear in the web function, however, these singularities appear additively, and require that all lines associated with the web become collinear at the same time.

 The collinear singularities of $E^{[1]}_{abr}$ associated with the incoming eikonals are to be removed by eikonal collinear subtractions, as in Eq.\ (\ref{eq:E-10}), and it is the subtracted form that must be analyzed for 
 renormalon or finite-mass power corrections.   The analysis is closely related to that for the Drell-Yan cross section \cite{Laenen:2000ij}.
 
 We next use the expansion of $\beta_r^\mu$ given in Eq.~(\ref{eq:beta-r-expand}) for the factor $\beta_r\cdot k$ in Eq.\ (\ref{eq:E-linear-gs}). The term proportional to $\beta_{r,T}\cdot k_T$ vanishes because the integrand is otherwise azimuthally symmetric.  It is then natural to combine the terms proportional to $\beta_a\cdot k$, and those proportional to $\beta_b\cdot k$, defining
\bea
G_a\ &\equiv&\ g_{ab} \ +\ g_{ar}\, ,
\nn\\[2mm]
G_b\ &\equiv&\ g_{ab} \ +\ g_{br}\, ,
\label{eq:G-ab-defs}
\eea
in terms of which (\ref{eq:E-linear-gs}) becomes,
\bea
E^{[1]}_{abr}(N/Q,\hat\eta,\mu,\epsilon) &=& -\ \frac{N}{Q}
\int {d^Dk\over (2\pi)^D}\ 
\left [\
{\beta_r\cdot \beta_a\, \beta_b\cdot k \over \beta_a\cdot \beta_b} \
G_a\left ( k^2,k_T^2 \right ) 
+ {\beta_r\cdot \beta_b\, \beta_a\cdot k \over \beta_a\cdot \beta_b}\
G_b\left ( k^2,k_T^2 \right ) \ \right ]
\nn\\[2mm]
&\equiv&\ {\cal E}_a^{[1]}(N/Q,\hat\eta,\mu,\epsilon) \ +\ {\cal E}_b^{[1]}(N/Q,\hat\eta,\mu,\epsilon) \, ,
\label{eq:E-linear-Gs}
\eea
where we exhibit only the momentum arguments for $G_a$ and $G_b$, and introduce a notation for their integrals.  All singularities for $k$ collinear to $\beta_a$ are in the ${\cal E}_a^{[1]}$ term, and all singularities for $k$ collinear to $\beta_b$ are in the 
${\cal E}_b^{[1]}$ term.

 We are now ready to  
analyze the $k$ integrals of $E^{[1]}_{abr}$, to all orders.    Because we are interested in singularities in the running coupling at small $\mu$, we limit ourselves to the region where $k_T^2 \le \kappa^2 \sim \Lambda^2_{\rm QCD}$.   An expansion in $\beta_r\cdot k$ is then similar to an expansion in energy for gluons emitted at low transverse momentum relative to the incoming directions.

\subsection{Infrared momenta and the absence of linear power corrections}

We recall that to identify the power corrections that emerge in the cross section, we have expanded 
the Laplace transform exponential, $\exp(-N\beta_r\cdot k/Q)$, in the exponent 
$E_{abr}$ of Eq.\ (\ref{eq:sig-abc}) in collinear and soft regions. This gives rise to the following conditions
on the energy of soft radiation:
\bea
k_0\, \le\, \frac{1}{\beta_a\cdot \beta_r} \frac{Q}{N}\, , \quad {\rm for} \ k\ {\rm collinear\ to}\ \beta_a\, ,
\nn\\[2mm]
k_0\, \le\, \frac{1}{\beta_b\cdot \beta_r} \frac{Q}{N}\, , \quad {\rm for} \ k\ {\rm collinear\ to}\ \beta_b\, .
\eea
In these terms, the eikonal cross section will depend on the overall scale $Q$ and the invariants $\beta_a\cdot \beta_r$ and $\beta_b\cdot \beta_r$.   
Correspondingly, in the subtraction terms in Eq.\ (\ref{eq:E-10}), we split the integral over $z$ 
into two regions,
\bea
\int_0^1 dz  \,\frac{ z^N-1}{1-z} \,=\,\int_0^{1-\frac{1}{N}} dz  \,\frac{ z^N-1}{1-z} \ + \ \int_{1-\frac{1}{N}}^1 dz  \,\frac{ z^N-1}{1-z} \, .
\label{eq:E-101}
\eea
The first region produces the familiar perturbative logarithm of $N$ in the moments of the anomalous dimension
for the eikonal parton distributions, whose role is to cancel a similar term in the eikonal hard-scattering functions
$\tilde \sigma^{\rm (eik)}_{abr}$~\cite{Laenen:2000ij}, as in Eq.\ (\ref{eq:tilde-omega}). We will not consider the integration over this region further. 
The second term, with the integration over $1-1/N\leq z\leq 1$, is relevant for our analysis of power corrections.
Making the approximation $z^N \sim \exp[-N(1-z)]$, which has corrections suppressed by additional powers of $N$,
and considering the region $k_T^2 \le \kappa^2 \sim \Lambda^2_{\rm QCD}$, we have
\bea
&&\hspace*{-10mm} \sum_{i=a,b}  \int_0^{\kappa^2} \frac{dk^2_T}{k_T^{2(1+\epsilon)}}\, A_i(\alpha_s(k_T^2)) \, \int_{1-\frac{1}{N}}^1 dz \, 
 \frac{{\mathrm{e}}^{-N(1-z)}-1}{1-z} \nn\\[2mm]
 & &\hspace{1cm}
 =\ - \sum_{i=a,b} \int_0^{\kappa^2} \frac{dk^2_T}{k_T^{2(1+\epsilon)}} \,A_i(\alpha_s(k_T^2)) \left[ 1-
 \sum_{P=2}^\infty \frac{(-1)^P}{P! \, P}\, \right ]\, .
\label{eq:subr-np}
\eea
This is a leading-power and collinear-singular correction, sensitive to low momentum scales,
 which will precisely cancel the singularities of the eikonal cross section, for $1-1/N<z<1$.
 We have isolated the first power, $P=1$ in the sum, which will cancel collinear singularities in $E_{abr}^{[1]}$, 
Eq.\ (\ref{eq:E-linear-Gs}), the first term in the eikonal cross section expanded in powers of $N\beta_r\cdot k/Q$.   We will return to the higher values of $P$ in the following subsection.

Eq.\ (\ref{eq:E-linear-Gs}) and the $P=1$ term of (\ref{eq:subr-np}) specify 
the full integral that includes the nonperturbative corrections that we are after.
Identifying these terms as they appear in Eq.~(\ref{eq:E-10}) we obtain
\bea
\hat E_{abr}^{[1]}(N/Q,\hat\eta,\mu)\ &=&\ 
{\cal E}_a^{[1]}(N/Q,\hat\eta,\mu,\epsilon) \ +\ \int_0^{\kappa^2} \frac{dk^2_T}{k_T^{2(1+\epsilon)}}\, A_a(\alpha_s(k_T^2)) \nn\\[2mm]
\ &+&\
{\cal E}_b^{[1]}(N/Q,\hat\eta,\mu,\epsilon) \ +\ \int_0^{\kappa^2} \frac{dk^2_T}{k_T^{2(1+\epsilon)}}\, A_b(\alpha_s(k_T^2))\, .
\label{eq:E1-sub}
\eea
We look first at the terms from the cross section that contain the collinear singularities in the $a$ direction,
again restricting to the range  $k_T^2 \le \kappa^2$: 
\bea
{\cal E}_a^{[1]}(N/Q,\hat\eta,\mu,\epsilon)\ &=&\
-\ \frac{N}{Q}\, \frac{\beta_a\cdot \beta_r}{\beta_a\cdot \beta_b}\,\Omega_{1-\epsilon}  \int_0^{\kappa^2} \frac{dk_T^2\, k_T^{-2\epsilon}}{2 (2\pi)^D}\ 
\int_0^{Q^2/(\beta_a\cdot \beta_r N)^2-k_T^2} 
dk^2\ G_a \left ( k^2,k_T^2 \right ) 
\nn\\[2mm]
&\ &\hspace{-32mm}  \times\
\int_{\sqrt{k^2+k_T^2}}^{Q/(\beta_a\cdot\beta_r N)} dk_0\ \int \frac{dk_3}{2|k_3|}\, \beta_b\cdot k\, \left [ \delta \left ( k_3 - \sqrt{k_0^2-k_T^2-k^2} \right )+ \delta \left ( k_3 + \sqrt{k_0^2-k_T^2-k^2} \right ) \right ]\,  \, ,
\nn\\
\label{eq:E-with-k0}
\eea
where $\Omega_{1-\epsilon}=2\pi^{1-\epsilon}/\Gamma(1-\epsilon)$ is the result of the angular integral that generalizes the azimuthal integral in dimensional regularization. Because of the boost-invariance of the web functions along the 3 axis, the energy integral can by carried out explicitly. With $\beta_b\cdot k=k_0+k_3$ we find
\bea
{\cal E}_a^{[1]}(N/Q,\hat\eta,\mu,\epsilon)\ &=&\
- \ \frac{N}{Q}\, \frac{\beta_a\cdot \beta_r}{\beta_a\cdot \beta_b}\,\Omega_{1-\epsilon}  \int_0^{\kappa^2} \frac{dk_T^2\, k_T^{-2\epsilon}}{2 (2\pi)^D}
\nn\\[2mm]
&\ &\hspace*{1.2mm} \times \int_0^{Q^2/(\beta_a\cdot \beta_r N)^2-k_T^2} dk^2\  G_a \left ( k^2,k_T^2 \right ) 
\, \sqrt{ \left ( \frac{Q}{\beta_a\cdot \beta_r N} \right)^2 - k^2 -k_T^2 }
\nn\\[2mm]
&=&\
-\frac{\Omega_{1-\epsilon}}{2\pi}   \int_0^{\kappa^2} {dk_T^2\, k_T^{-2\epsilon}}
\int_0^{Q^2/(\beta_a\cdot \beta_r N)^2-k_T^2} \frac{dk^2}{4(2\pi)^{D-1}}\  G_a \left ( k^2,k_T^2 \right ) 
\nn\\[2mm]
&\ &\hspace*{1.2mm} \times
 \left [ 1  - \frac{ \left(k^2 +k_T^2\right ) (\beta_a\cdot \beta_r)^2 N^2}{2Q^2} + \dots \right ]\, ,
\label{eq:E-with-k0-done}
\eea
where in the second equality, we have used $\beta_a\cdot \beta_b=2$, 
in the normalization of Eq.\ (\ref{eq:beta-ab}).
In the expansion, we identify a leading-power term, corrections
beginning at the $N^2/Q^2$ level, as anticipated.   
As discussed after Eq.\ (\ref{eq:E-10}), factorization ensures that the singularities associated with the leading-power term 
from the web function integrals are cancelled by the 
subtraction in Eq.\ (\ref{eq:E1-sub}) from the eikonal parton distribution \cite{Laenen:2000ij}.   
This implies that
\bea
\int_0^{Q^2/(\beta_a\cdot \beta_r N)^2-k_T^2} \frac{dk^2}{4(2\pi)^{D-1}}\  G_a \left ( k^2,k_T^2 \right ) \ =\ \frac{A_a\left (\alpha_s(k_T^2)\right)}{k_T^2}\ +\ 
{\cal A}_a(k_T^2,Q^2)\, ,
\label{eq:Ga-to-A}
\eea
where the function ${\cal A}_a$ is collinear finite.   In fact, because ${\cal A}_a$ can depend only on $k_T^2$ and $Q^2$, it must be suppressed 
by a power of $Q/N$. More importantly, by the properties of web functions, it has no logarithmic scales except for $Q/(N\mu)$.   
The scale of the running coupling is thus naturally chosen as $Q/N$ in this term.  
This is still a perturbative scale, and does not imply a nonperturbative power correction from the integral over $k_T$. 
We assume, therefore, that it can be neglected.  
A result completely analogous to Eq. (\ref{eq:Ga-to-A}) holds for collinear radiation in the $b$ direction, with subscript $b$ everywhere. 
Since collinear singularities now cancel, we can return our
full expression in Eq.~(\ref{eq:E1-sub}) to $D=4$.

The leading correction that remains after the cancellation of collinear singularities is then 
obtained from (\ref{eq:E1-sub}) as (we recall from~\ref{eq:G-ab-defs}) that the flavor index $r$ for the recoil parton is implicit
in the functions $G_{a,b}$):
\bea
\hat E^{[1]}_{abr} (N/Q,\hat\eta,\mu)  \ &=&\  \frac{  (\beta_a\cdot \beta_r)^2 N^2}{2Q^2}
  \int_0^{\kappa^2} {dk_T^2\, }
\int_0^{Q^2/(\beta_a\cdot \beta_r N)^2-k_T^2} \frac{dk^2}{4(2\pi)^{3}}\  \left(k^2 +k_T^2\right )\ 
\nn\\[2mm]
&\ & \hspace{26mm} \times\ \left [ G_a \left ( k^2,k_T^2,\alpha_s(k_T^2) \right ) + G_b \left ( k^2,k_T^2, \alpha_s(k_T^2) \right ) \right ]
\,   \, .
\label{eq:1-over-Q2}
\eea
In this expression, we have reintroduced dependence on the running coupling in the functions $G_{a,b}$, using 
their renormalization-scale independence, Eq.\ (\ref{eq:g-rg}).   We note that by the general properties of the web functions discussed in Sec.\ \ref{sec:no-subs},
all logarithmic behavior is associated with $k^2$ and $k_T^2$ vanishing at the same time.   Then, choosing $\mu^2=k_T^2$ absorbs all
perturbative logarithms into the running coupling.   The ambiguities in the low-scale behavior of the coupling induce the
power corrections to perturbation theory.

On the basis of this discussion, we see that the term nominally power-suppressed by $N/Q$ includes a leading power contribution from values of $k_0$ of order
$Q/N$, which, however, is cancelled by the collinear subtraction.   Nevertheless, the next power corrections 
appear at order $N^2/Q^2$, with no linear, $N/Q$ contribution.
To complete our argument on the absence of linear power corrections, we return now to higher orders in the expansion in $N\beta_r\cdot k/Q$.

\subsection{Higher orders in the power expansion}

We know that higher powers in the expansion of the exponent Eq.\ (\ref{eq:E-10}) contribute at leading and nonleading powers in $N/Q$, because they are necessary to match to the full collinear subtraction, beyond the first term in the expansion shown in Eq.\ (\ref{eq:subr-np}).  

The treatment of these terms closely follows that of the lowest order, $N/Q$, just described in detail. We consider order $P$ in the expansion of the exponent, and note that the arguments of Sec.\ \ref{sec:lo-beyond} apply to arbitrary powers of 
the weight $\beta_r\cdot k$.  The general form is (see Eq.~(\ref{eq:E-1})) 
\be
E^{[P]}_{abr}(N/Q,\hat\eta,\mu,\epsilon)\ \equiv\ -\ \frac{1}{P!} \left ( \frac{N}{Q} \right )^P 
\int {d^Dk\over (2\pi)^D}\ \left(\beta_r\cdot k \right)^P 
\rho_{abr}^{[P]}
\left(\left \{ {\beta_i\cdot k\,  \beta_j\cdot k \over \beta_i\cdot\beta_j } \right \},k^2,\mu^2,
\alpha_s(\mu^2)\right)\, ,
\label{eq:E-P}
\ee
where $\rho_{abr}^{[P]}$ is the web function that remains after the cancellation of final state interactions.   Generalizing the case of $P=1$, the product $(\beta_k\cdot kj)^P \rho_{abr}^{[P]}$ is proportional to the $P$th power of $\beta_r$ and of dot products of $k$ and the $\beta_i$, $i=a,b,r$, and is otherwise a function of boost-invariant 
factors of $k^2$ and $k_T^2$, using the frame chosen above. We may thus write
\bea
\left(\beta_r\cdot k \right)^P\, 
\rho_{abr}^{[P]}
\ =\ \sum_{A=0}^P \sum_{B=0}^{P-A}\, \left( \beta_r\cdot k \right)^{P-A-B} \, \left ( {\beta_r\cdot \beta_a\, \beta_b\cdot k \over \beta_a\cdot \beta_b} \right)^A
\left ( {\beta_r\cdot \beta_b\, \beta_a\cdot k \over \beta_a\cdot \beta_b} \right)^B\ g^{(P)}_{AB}(k^2,k_T^2)\, ,
\label{eq:gAB-sum}
\eea
where by comparison to the $P=1$ case, (\ref{eq:linear-beta-c}) we can identify
\bea
g^{(1)}_{00} \ &=&\ g_{ab}\, ,
\nn\\[2mm]
g^{(1)}_{10} \ &=&\ g_{ar}\, ,
\nn\\[2mm]
 g^{(1)}_{01} \ &=&\ g_{br}\, .
 \eea
 Expanding factors of $\beta_r\cdot k$ as in Eq.\ (\ref{eq:beta-r-expand}), we derive a set of terms that are polynomials in $\beta_a\cdot k$, 
 $\beta_b\cdot k$, and in even powers of 
 $\beta_{r,T}\cdot k_T$.\footnote{We have simplified Eq.\ (\ref{eq:gAB-sum} somewhat by neglecting terms with invariants $\beta_r\cdot \beta_a \beta_r\cdot \beta_b/\beta_a\cdot \beta_b$ times $k^2$ or $k_T^2$.  In such terms, the leading contributions begin at order $1/Q^2$.}
 
 At $P$th order in the power expansion, we want to identify terms in the integral over $k$ that can compensate for the overall power of $(N/Q)^P$ in Eq.\ (\ref{eq:E-P}), and to confirm that there are no corrections at linear, $1/Q$, order.   As in the case of $P=1$, contributions that grow with $Q$ can come only from the $k_0$ integral.   Although for large $P$ there are many
terms in Eq.\ (\ref{eq:gAB-sum}), powers $Q^0$ and $Q^{-1}$ can come about only from factors $(\beta_b\cdot k)^P$, which give collinear singularities in the $\beta_a$ direction, or $(\beta_a\cdot k)^P$, which give collinear singularities in the $\beta_b$ direction.   In particular, any factor $\beta_a\cdot k \beta_b\cdot k=(k^2+k_T^2)/2$ or  
 $(\beta_{r,T}\cdot k_T)^2$ gives corrections that suppress the integral by $1/Q^2$.   
 
In summary, the only possible sources of $1/Q$ corrections are integrals of the type
\bea
{\cal E}_a^{[P]}(N/Q,\hat\eta,\mu,\epsilon)\ &=&\ -\
\frac{1}{P!}\, \left ( \frac{N}{Q} \right )^P 
\int {d^Dk\over (2\pi)^D}\ \left( \frac{\beta_r\cdot \beta_a \beta_b\cdot k }{ \beta_a\cdot \beta_b}\right)^P\ G_a^{[P]} \left(k^2,k_T^2\right)\, ,
\label{eq:E-P-def}
\eea
where $G_a^{[P]}$ generalizes $G_a$ in Eq.\ (\ref{eq:G-ab-defs}),
\bea
G_a^{[P]}\ =\ \sum_{A=0}^P \sum_{B=0}^{P-A}\, g^{[P]}_{AB} \left ( k^2,k_T^2 \right)\, .
\label{eq:G-P-def}
\eea
Expanding the integrals, this is
\bea
{\cal E}_a^{[P]}(N/Q,\hat\eta,\mu,\epsilon)\ &=&\ - \ \frac{1}{P!}
 \left( \frac{N}{Q} \right )^P\, \left( \frac{\beta_a\cdot \beta_r}{\beta_a\cdot \beta_b} \right)^P\,\Omega_{1-\epsilon}  \int_0^{\kappa^2} \frac{dk_T^2\, k_T^{-2\epsilon}}{2 (2\pi)^D}\nn\\[2mm]
&\ & \hspace*{-1.5cm} \times  \int_0^{Q^2/(\beta_a\cdot \beta_r N)^2-k_T^2} dk^2\ G_a^{[P]} \left ( k^2,k_T^2 \right ) 
\int_{\sqrt{k^2+k_T^2}}^{Q/(\beta_a\cdot\beta_r N)} dk_0\, 
\nn\\[2mm]
&\ &\hspace*{-1.5cm} \times\  \int \frac{dk_3}{2|k_3|}
\left( \beta_b\cdot k \right)^P \, \left [ \delta \left ( k_3 - \sqrt{k_0^2-k_T^2-k^2} \right )+ \delta \left ( k_3 + \sqrt{k_0^2-k_T^2-k^2} \right ) \right ]\,  \, .
\nn\\
\label{eq:EP-with-k0}
\eea
It is not difficult to show that 
although this integral produces exactly the leading power in the $P$th terms of the collinear subtraction Eq.\ (\ref{eq:subr-np}), 
its corrections linear in $N/Q$ vanish, leaving corrections that begin at $(N/Q)^2$.  To verify the this claim, we can consider the class of integrals,
\bea
I_{m,n}(Q/N)\ \equiv \ \left(\frac{N}{Q}\right )^{m+n+1} \ \int_{l_0}^{Q/N} dk_0\, k_0^m\, \left( k_0^2 - l_0^2 \right)^{n/2}\, ,
\eea
with $l_0^2=k^2+k_T^2$. The case $P=1$ is realized by $m=1$, $n=-1$ or by $m=n=0$, but only the former is relevant here
since for $m=n=0$ the $k_3$ integral in Eq.\ (\ref{eq:EP-with-k0}) vanishes by antisymmetry of the integrand. 
The $P>1$ integrals provide $m+n \ge 1$.   Adding and subtracting $k_0^{m+n}$ in the integrand, and changing variables to $x=k_0/l_0$, we find
\bea
I_{m,n}(Q/N)\ &=&\ \left(\frac{N}{Q}\right )^{m+n+1} \, \left \{ \frac{1}{m+n+1}  \left( \left(\frac{Q}{N}\right)^{m+n+1} - \, l_0^{m+n+1} \right )
\right.
\nn\\[2mm]
&- &\left.\ l_0^{m+n+1}\
\int_1^{Q/(Nl_0)} dx \, x^{m+n} 
\left( 1 - \left( 1- \frac{1}{x^2} \right)^{n/2} \right )\right \}
\nn\\[2mm]
&=& \hspace{5mm} \ \frac{1}{m+n+1} \ +\ {\cal O}\left ( \frac{N^2}{Q^2} \right )\, \, .
\eea
Substituting this result into Eq.\ (\ref{eq:EP-with-k0}),  the $Q^0$ contribution leads to a collinear-singular integral over $k_T^2$.  
This is just that part of the standard initial-state collinear singularity of the eikonal cross section that shows up in the $(N/Q)^P$ term of the expansion of the exponential in Eq.\ (\ref{eq:E-10}).  As such, the collinear finiteness of the eikonal hard-scattering function $\tilde\omega_{abr}^{({\rm eik})}$ again ensures that the dimensional poles of this integral will
match the corresponding term in the expansion of the eikonal collinear subtraction, Eq.\ (\ref{eq:subr-np}). This implies a relation between the function $G_a^{[P]}$ and the anomalous dimension $A_a(\alpha_s)$, analogous to Eq.\ (\ref{eq:Ga-to-A}).

It will be of interest for future work to study the systematics of the full set of $(N/Q)^2$ nonperturbative contributions, in terms of their low-order expansions in the running coupling, using methods related to those in Refs.\ \cite{Korchemsky:1994is} and \cite{Dokshitzer:1995qm},
and perhaps 
the all-order structure of these contributions.

\section{Conclusions and outlook}

We have shown in full QCD the absence of coupling-induced linear power corrections in the eikonal approximation, for electroweak boson production at fixed transverse momentum and rapdity. This result is consistent with the recent detailed analysis at order $g^2$ based on models with a massive gluon \cite{Caola:2021kzt} 
and in particular the use of azimuthal symmetry for isolated terms proportional to a single power of  $k_T$.    
It is worth noting that the set of integrals considered in Ref. \cite{Caola:2021kzt} includes several that reflect the recoil of hard partons to soft radiation.
These terms are not included in our analysis because they are suppressed by a power of the soft gluon energy, and hence by $1/N$ in the Laplace moment, before expansion of the exponential.  Such terms are independent of the resummation of threshold logarithms.
 
Looking ahead, it may be possible to generalize our analysis beyond the eikonal approximation,  relying on the 
``next to eikonal" structure of these cross sections investigated in recent years 
\cite{Laenen:2010uz,Bonocore:2016awd,DelDuca:2017twk,Bahjat-Abbas:2019fqa}.  
We also expect our results to have valuable ramifications for phenomenology, given the
vast body of experimental measurements for single electroweak boson production that have been carried out. 
Prompt photons produced in fixed-target scattering have presented a long-standing challenge to 
theory \cite{FermilabE706:1997naa,Baur:2000xd,Aurenche:2006vj}, which makes an improved
treatment of power corrections highly desirable. Similarly, recent studies \cite{Gonzalez-Hernandez:2018ipj,Bacchetta:2019tcu}
have found that fixed-order and even threshold-resummed perturbation theory for semi-inclusive deep-inelastic scattering 
and the Drell-Yan process at large transverse momentum do not describe the
available experimental data well in the fixed-target regime. Equipped with a more robust understanding
of power corrections, one may now attempt to resolve the discrepancies between data and theory for
these reactions, perhaps using models of the strong coupling, as for example, in Ref.\ \cite{Dokshitzer:1995qm}.

Finally, it will clearly be very interesting to extend our study to the case of ``pure QCD'' 
processes, that is, replacing the produced electroweak boson by a QCD parton that subsequently fragments
into a hadron or produces a jet. In this way, one can address power corrections for a larger
class of single-particle reactions, where the eikonal approximation involves the full complexity
of QCD. Here the question as to whether power corrections start at order $1/p_T$ or $1/p_T^2$
is still open. Its resolution will likely impact the description of fixed-target scattering data for
single-hadron production, as well as the fragmentation contribution to the prompt-photon cross
section. In this context, it is interesting to note that recent Phenix photon data~\cite{PHENIX:2022lgn}
are in better agreement with NLO theory for the isolated case than for the non-isolated one at
$p_T\lesssim 10$~GeV, the latter having a much larger fragmentation component.

\subsection*{Acknowledgements}  The work of GS was supported in part by the National Science Foundation, award PHY-1915093.
WV is grateful for support by the Bundesministerium f\"{u}r Bildung und Forschung (BMBF), grant 05P21VTCAA.

\end{document}